\documentclass[11pt,notitlepage]{article}

\usepackage[margin=1.0in]{geometry}

\usepackage[square,sort,comma,numbers]{natbib}

\usepackage[linesnumbered,ruled]{algorithm2e} % for algorithms environment
\usepackage{colortbl}  % for black cells in matrices
\usepackage{makecell}  % for word wrapping in table 
\usepackage[normalem]{ulem} 
\usepackage{subcaption}
\usepackage{hyperref}
\usepackage{pifont}
\usepackage{tikz}
\usetikzlibrary{calc}      % to calculate coordinates
\usetikzlibrary{patterns}  % for cross hatching

\usepackage{environ}
\NewEnviron{tikscale}{
    \begin{center}
    \scalebox{0.7}{
        \begin{tikzpicture}
            \BODY
        \end{tikzpicture}
    }
    \end{center}
}

\usepackage{fancyhdr}
\usepackage{amsmath}
\usepackage{amsfonts}
\usepackage{amssymb}
\usepackage{amsthm}
\theoremstyle{definition}
\newtheorem{definition}{Definition}[section]
\newtheorem{theorem}{Theorem}

\newcommand{\Rs}{\textbf{r}}
\newcommand{\Cs}{\textbf{c}}
\newcommand{\arc}{\mathcal{A}(\Rs,\Cs)}

\newcommand{\arcp}{\mathcal{A'}(\Rs,\Cs)}
\newcommand{\W}{W}
\newcommand{\A}{A}
\newcommand{\B}{B}
\newcommand{\CC}{C}
\newcommand{\D}{D}
\newcommand{\E}{E}
\newcommand{\F}{F}
\renewcommand{\P}{\mathbb{P}}
\newcommand{\sz}{\cellcolor[rgb]{0.6,0.6,0.6}}

\providecommand{\keywords}[1]
{
  \small	
  \textbf{\textit{Keywords---}} #1
}

\begin{document}

\title{Non-Uniform Sampling of Fixed Margin Binary Matrices}

\author{Alex Fout \\ fout@colostate.edu \\ Department of Statistics \\ Colorado State University
\and Bailey K. Fosdick \\ bailey.fosdick@colostate.edu \\ Department of Statistics \\ Colorado State University
\and Matthew P. Hitt \\ matthew.hitt@colostate.edu \\ Department of Political Science \\ Colorado State University}

\maketitle

\begin{abstract}
    Data sets in the form of binary matrices are ubiquitous across scientific domains, and researchers are often interested in identifying and quantifying noteworthy structure. 
    One approach is to compare the observed data to that which might be obtained under a null model.
    Here we consider sampling from the space of binary matrices which satisfy a set of marginal row and column sums.
    Whereas existing sampling methods have focused on uniform sampling from this space, we introduce modified versions of two elementwise swapping algorithms which sample according to a non-uniform probability distribution defined by a weight matrix, which gives the relative probability of a one for each entry.
    We demonstrate that values of zero in the weight matrix, i.e. \emph{structural zeros}, are generally problematic for swapping algorithms, except when they have special monotonic structure.
    We explore the properties of our algorithms through simulation studies, and illustrate the potential impact of employing a non-uniform null model using a classic bird habitation dataset.
\end{abstract}


\section{Introduction}

In 1975, Diamond~\cite{diamond_assembly_1975} proposed rules which govern the combination of several bird communities on islands in New Guinea, such as ``Some pairs of species never coexist, either by themselves or as a part of a larger combination''.
However, Connor \& Simberloff~\cite{connor_assembly_1979} disputed these rules, suggesting that the combination of species on the islands was not exceptionally different from random. 
Aside from the ensuing feud within the ecological community~\cite{diamond_examination_1982,besag_generalized_1989,stone_checkerboard_1990,manly_note_1995,sanderson_null_1998,gotelli_null_2000,gotelli_swap_2001}, Connor \& Simberloff sparked interest in a novel form of ``null model analysis''~\cite{harvey_null_1983,gotelli_null_1996}, a general approach which compares an observed matrix statistic to the distribution of that statistic under a null model. 
The null model of choice in this case was the set of all random combinations of species on islands satisfying the following three constraints:
\begin{itemize}
    \item The number of species on each island matches the number observed.
    \item The number of islands inhabited by each species matches the number observed.
    \item Each species inhabits islands of comparable size to the islands on which it is observed. 
\end{itemize}
Connor \& Simberloff reason that if the observed species combination is not substantially different from the random combinations with respect to some statistic of interest, then it becomes difficult to argue that the observed pattern is anything noteworthy.

To generate combinations which adhere to the above constraints, they represent the species/~island combinations as a matrix, with rows representing species and columns representing islands. 
The element in a particular row and column is set to 1 if the corresponding species inhabits the corresponding island, and 0 otherwise. 
In the matrix representation, the first two constraints imply fixing the marginal sums of the matrix, while the third constraint limits which elements are permitted to be 1.
Thus, the aim in the ``null model'' analysis is to generate a collection of matrices which possess the desired marginal sums and exclude the prohibited 1's.
Connor \& Simberloff propose two methods, one which begins with a 0-filled matrix and sequentially \textit{fills} in each row randomly, being careful to adhere to the constraints, and another which begins with the observed matrix, and repeatedly \textit{swaps} 1's such that the constraints are maintained.
Subsequent authors have proposed additional methods for generating the desired matrices, which are generally either a ``fill''~\cite{wilson_methods_1987,stone_checkerboard_1990,sanderson_null_1998,gotelli_swap_2001} or ``swap''~\cite{besag_generalized_1989,roberts_island-sharing_1990,stone_checkerboard_1990,manly_note_1995,zaman_random_2002} type method.

Early debates surrounding the use of this particular null model focused on the appropriateness of imposing Connor \& Simberloff's constraints, the choice of  statistic upon which to base comparison, and the interpretability of null model tests~\cite{diamond_examination_1982,roberts_island-sharing_1990}.  
However, focus eventually shifted from whether to use this approach to which method of producing matrices is best~\cite{sanderson_null_1998}.
This accompanied theoretical work in binary matrices with fixed marginal sums~\cite{brualdi_matrices_1980} and Markov Chain Monte Carlo (MCMC) sampling~\cite{besag_generalized_1989},  paving the way for swap and fill methods with proven statistical properties~\cite{kannan_simple_1999,chen_sequential_2005}.

Generally speaking, sampling from a null model can be useful for quantifying the degree of ``non-randomness'' in an observed matrix (although care should be exercised in the selection of a test statistic and mechanistic interpretation of results). 
This approach has been used for testing for significance of species co-occurrence~\cite{connor_checkered_2013}, evaluating goodness of fit for the Rasch item-response model~\cite{besag_generalized_1989,chen_sequential_2005}, and in the analysis of contingency tables~\cite{chen_sequential_2005}.

In this paper, we extend the capability of null model analysis by allowing for non-uniform sampling.
This permits the construction of more sophisticated ``null'' models corresponding to non-uniform probability distributions.

\subsection{Review of Sampling Methods}
Here we briefly review the variety of existing swap and fill methods that have been employed for sampling binary matrices with fixed margins. 
To highlight the contribution of this work, we point out that most efforts at sampling these matrices thus far have focused on uniform sampling, where each matrix has an equal probability of being included in the sample.
Early work gave little attention to actually proving uniformity and few authors address situations in which some matrix elements must be 0, so called \textit{structural zeros}, with exceptions noted below.

As mentioned, these sampling methods can be categorized into \textit{fill} and \textit{swap} methods.
Fill methods are typically faster than swap methods, but risk hitting ``dead ends'', where the process must restart or backtrack in order to avoid violating row and column margin constraints.
Swap methods, on the other hand, are arguably simpler to understand and implement, but rely on local changes, so sampled  matrices tend to be highly correlated.
For both categories, initial methods lacked much theoretical treatment, but more recent work has considered the statistical properties of the algorithms. 

\subsubsection{Fill Methods}
Fill methods begin with a matrix of all 0's, and randomly \textit{fill} in 1's in such a way that adheres to the margin (row and column) constraints.

Connor \& Simberloff's~\cite{connor_assembly_1979} first placed the row sums in decreasing order and column sums in increasing order (``canonical form''), then proceeded row by row, randomly selecting columns to fill with 1's, until the desired row sum is reached. 
If the column is already 1, or it has met its column sum, another random selection is made. 
If the procedure reaches a point where no 1 can be filled without violating some constraint (a dead end), the process restarts with an empty matrix.
No mention is made of the probability of generating a particular matrix. In their ``Milne method'', Stone \& Roberts~\cite{stone_checkerboard_1990} adapt the fill method of Connor \& Simberloff to check whether the proposed fill will lead to a dead end.

Sanderson et al~\cite{sanderson_null_1998} propose a ``knight's tour'' algorithm which randomly samples a row and column index for filling, rather than filling one row at a time, and backtracks rather than restarts if the algorithm reaches a dead end. 
They claim that since row and columns are sampled randomly, then each valid matrix occurs with equal probability.
However, Gotelli \& Entsminger show that the ``knight's tour'' algorithm does not sample uniformly, and propose a ``random knight's tour'' which only samples a subset of available elements before backtracking, and backtracks randomly rather than sequentially. 
They provide no proof of uniform sampling, instead showing on a small example that a sample generated by the random knight's tour does not reject a lack of fit test.

Chen et al~\cite{chen_sequential_2005} propose a sequential importance sampling approach which randomly fills in each row, according to a conditional Poisson proposal distribution that depends on the previously filled rows. 
The matrices in the sample are then re-weighted to approximate a uniform distribution.
This method also approximates the size of the space of valid matrices.

Taking a different approach, Miller \& Harrison~\cite{miller_exact_2013} devise a recursive sampling algorithm based on the Gale-Ryser Theorem to generate matrices from the uniform distribution exactly, and compute the size of the space exactly.
The recursion is over row and column sums, so this method is limited to matrices of moderate size.

Finally, Harrison \& Miller~\cite{harrison_importance_2013} adopt the sequential importance sampling approach of Chen et al for non-uniform sampling. 
This is, to our knowledge, the only paper dedicated towards deliberate non-uniform sampling. 
They propose a multiplicative weights model that forms the basis of our weights model below, and permit specifying structural zeros by setting a corresponding weight to 0.

\subsubsection{Swap Methods}
Swap methods have developed in parallel to fill methods, beginning with the original paper of Connor \& Simberloff~\cite{connor_assembly_1979}.
These methods involve identifying ``checkerboard'' patterns of 0's and 1's in an existing valid matrix, and swapping the 1's for the 0's to produce a mew matrix which still adheres to the constraints (see Figure~\ref{fig:cb_swap}).
Repeated swaps result in a Markov chain Monte Carlo algorithm that produces a dependent sample of valid matrices. 
Of particular importance is the method used to identify potential swaps, and what happens if a proposed swap is invalid, which have computational as well as theoretical implications.
Initially, there was some question about whether swapping methods could produce all possible valid matrices by starting at the observed matrix; in other words, if the space of valid matrices is connected under checkerboard swaps.
However, Brualdi~\cite{brualdi_matrices_1980} showed this to be true using elementary circuits. 
This paved the way for more development in swap methods~\cite{stone_checkerboard_1990,manly_note_1995}.

%Connor \& Simberloff~\cite{connor_assembly_1979} scan the matrix for checkerboard patterns in order to identify a potential swap. 
%After several swaps, the new matrix is recorded if it is unique from the previous matrix when placed in canonical form.
%No detail is provided about the scanning procedure or the number of swaps between samples, and little discussion is given to the impact of structural zeros.

%Stone \& Roberts~\cite{stone_checkerboard_1990} invoke Brualdi to prove that their swapping method can reach all valid matrices, but leave unanswered the question of whether the sample is uniform.
%Again, there is ambiguity about the precise method of identifying checkerboard swaps, which turns out to be critical in determining whether the sampling process is uniform.

%Manly~\cite{manly_note_1995} proposes a swapping procedure which envisions the observed matrix as randomly occurring somewhere in a swap sequence of given length. Therefore, starting from the observed matrix, they perform ``forward'' swaps, and ``backward'' swaps to create a swap sequence which contains the observed matrix. They then invoke theory from Besag \& Clifford~\cite{besag_generalized_1989} to perform exact hypothesis testing against all sequences of the given length containing the observed matrix.
%Importantly, this is not the same as testing against all matrices satisfying the constraints.

As with fill methods, early swap methods gave no proof that matrices are sampled uniformly, until Kannan et al~\cite{kannan_simple_1999} proved this for the following procedure:
Select two rows and two columns uniformly at random.
If they form a checkerboard, then perform a swap, and if not, keep the previous matrix.
This result influenced the development of subsequent swapping algorithms.
Gotelli \& Entsminger's~\cite{gotelli_swap_2001} version of the swapping algorithm importantly picks a swap randomly from the set of eligible swaps with equal probability, not as prescribed by Kannan et al, and consequently demonstrate that the resulting distribution which this samples from is not uniform.
This approach is also taken by Zaman \& Simberloff~\cite{zaman_random_2002}, who correct for the imbalance by reweighting each matrix when computing summary statistics.
These approaches require identifying all possible swaps at each step in order to sample uniformly from them, which is prohibitive for large matrices.

Finally, in the ``curveball'' algorithm proposed by Strona et al~\cite{strona_fast_2014} multiple swaps are performed at once by first selecting two rows (or columns), and permuting the elements in their symmetric difference.
Following the proof technique of Kannan et al., Carstens~\cite{carstens_proof_2015} proves that curveball samples uniformly and demonstrate that multiple swaps can lead to faster mixing compared to Kannan et al's algorithm.

\subsection{Non-Uniform Sampling}
Most existing null matrix sampling algorithms focus on the uniform distribution, the goal being to compare the observed matrix to other matrices with the same marginal properties, without preference for particular matrices over others.
We consider, however, that for some data analyses we may wish to lend more importance to some matrices compared to others, when testing for a specific effect.
For instance, Diamond~\cite{diamond_examination_1982} mentions several factors which may give rise to certain species combinations, such as resource overlap, dispersal ability, proneness to local extinction, and distributional strategy,  and criticised Connor \& Simberloff~\cite{connor_assembly_1979} for not accounting for such ``significant structuring forces''.
We argue that it may be possible to control for such external effects in a modified null model, which does not give equal probability to all possible matrices.

Here we consider the task of non-uniform sampling of matrices in a general setting.
Following the formulation of Harrison \& Miller~\cite{harrison_importance_2013}, we assume the existence of a weight matrix indicating the relative likelihood of a 1 in each element and define the probability of a matrix as a function of these weights. 
Careful attention must be paid to the consequences of weights equal to 0, which further restrict the set of matrices with positive probability, and affect the validity of the proposed sampling methods.

In the rest of this paper, we describe a weight-based probability model, then present a novel MCMC checkerboard swap algorithm which incorporates the weights and samples according to the given model. 
The weights introduce the notion of a \emph{structural zero}, which can impact the efficiency and even correctness of sampling.
We then present a novel curveball MCMC algorithm, which in most circumstances is a faster alternative to the weighted checkerboard swap algorithm.
We investigate the correctness, efficiency under different weighting and structural zero schemes %, and sensitivity to relative weight magnitudes of these algorithms
via simulation.
Finally, we revisit the bird species-island dataset first discussed by Diamond~\cite{diamond_assembly_1975} and Connor \& Simberloff~\cite{connor_assembly_1979} and consider incorporating a weight matrix in the null distribution to show the impact that non-uniform sampling can have on scientific conclusions.

\section{Non-Uniform Sampling from binary Matrices}

\subsection{Probability Model}

Let $\arc$ be the set of all $m \times n$ binary matrices satisfying row sums $\Rs=(r_1, r_2, \ldots, r_m)$ and column sums $\Cs=(c_1, c_2, \ldots, c_n)$. Furthermore let $\W = (w_{ij}) \in [0, \infty)^{m\times n}$ be a non-negative weight matrix representing the relative likelihood of a 1 in element $(i, j)$. 
Following Harrison \& Miller~\cite{harrison_importance_2013}, define the probability of observed matrix $\A = (a_{ij}) \in \arc$ as 
\begin{align}
\P(\A) &= \frac{1}{\kappa} \prod_{ij} w_{ij}^{a_{ij}},\quad \kappa = \sum_{\A \in \arc} \prod_{ij} w_{ij}^{a_{ij}}.
\label{eq:model}
\end{align}
We call the set of elements $\{(i,j) : w_{ij}=0\}$ \textit{structural zeros}, because any matrix with positive probability must have $a_{ij}=0$ if $w_{ij}=0$.
Let $\arcp = \{\A \in \arc : \P(\A)>0\}$ be the set of matrices with positive probability.
We interpret the weights by considering the relative probability between two matrices $\A, \B \in \arcp$:
\begin{align}
    \frac{\P(\A)}{\P(\B)} &= %\frac{\displaystyle \prod_{ij: a_{ij}=1} w_{ij}}{\displaystyle \prod_{ij: b_{ij}=1} w_{ij}} = 
    \frac{\displaystyle \prod_{ij: \substack{a_{ij}=1\\ b_{ij}=0}} w_{ij}}{\displaystyle \prod_{ij: \substack{b_{ij}=1\\ a_{ij}=0}} w_{ij}}, \label{eq:PBPA}
\end{align}
which is governed by the weights of the elements not shared between them.
Furthermore, we can write the conditional probability:
\begin{align}
    %\P(\B | \{\A , \B \}) & 
    \P(\B | \{\A \} \cup \{\B \}) & 
    = \frac{\displaystyle \prod_{ij: \substack{b_{ij}=1 \\ a_{ij}=0}} w_{ij}}{\displaystyle \prod_{ij: \substack{a_{ij}=1 \\ b_{ij}=0}} w_{ij} + \prod_{ij:\substack{b_{ij}=1 \\ a_{ij}=0}} w_{ij}}.
\end{align}
This forms the basis for the swapping probability defined in the next section.

\subsection{Weighted Checkerboard Swap Algorithm}
\label{sec:wcs}

\begin{figure}
    \centering
    \begin{tabular}{c|c|c|}
    \multicolumn{1}{c}{} & \multicolumn{1}{c}{$j$} & \multicolumn{1}{c}{$j'$} \\ \cline{2-3}
        $i$ & 1 & 0 \\\cline{2-3}
        $i'$& 0 & 1 \\\cline{2-3}
    \end{tabular}
    \quad
    \quad
    $\longrightarrow$
    \quad
    \quad
    \begin{tabular}{c|c|c|}
    \multicolumn{1}{c}{} & \multicolumn{1}{c}{$j$} & \multicolumn{1}{c}{$j'$} \\ \cline{2-3}
        $i$ & 0 & 1 \\\cline{2-3}
        $i'$& 1 & 0 \\\cline{2-3}
    \end{tabular}
    \caption{A $2 \times 2$ submatrix of $\A$ from rows $i$ and $i'$, and columns $j$ and $j'$, which are not necessarily consecutive. Swapping the positions of 1's and 0's preserves the row and column sums of $\A$.}
    \label{fig:cb_swap}
\end{figure}

We propose a swapping algorithm similar to that of Brualdi~\cite{brualdi_matrices_1980}, which identifies $2 \times 2$ submatrices with diagonal 1's and 0's, as in Figure~\ref{fig:cb_swap}, where the rows and columns are not necessarily consecutive.
Let $(i, j)$ and $(i', j')$ be the indices of the 1's in the submatrix, so that $(i, j')$ and $(i', j)$ are 0's.
The positions of the 1's and 0's can be swapped without altering the row or column sums.
This swap operation produces a matrix $\B \in \arc$ which differs from $\A$ (see Figure~\ref{fig:cb_swap}).
In the uniform sampling case, all matrices have equal probability, so swaps are performed with a fixed probability.
In our version, the swap is performed with probability
\begin{equation}
%\P(\B | \{\A, \B \}) = \frac{w_{i'j}w_{ij'}}{w_{ij}w_{i'j'} + w_{i'j}w_{ij'}}:= p_{ij;i'j'}.
\P(\B | \{\A \} \cup \{\B \}) = \frac{w_{i'j}w_{ij'}}{w_{ij}w_{i'j'} + w_{i'j}w_{ij'}}:= p_{ij;i'j'}.
\end{equation}
%and staying at $\A$ otherwise.
This modification implies matrices with zero probability will be visited with probability zero.

%The weighted checkerboard swap forms the basis of an MCMC sampling algorithm which samples from $\arcp$ according to the probability model in \eqref{eq:model}, described in Algorithm \ref{alg:weighted_swap}.

In Algorithm~\ref{alg:weighted_swap}, we present an MCMC procedure based on weighted checkerboard swap. % to sample from $\arcp$ according to the probability model in \eqref{eq:model}.
As with any MCMC algorithm, this algorithm can be modified to incorporate burn-in and thinning to reduce autocorrelation between saved samples.

\begin{algorithm}
\caption{Sampling Via Weighted Checkerboard Swaps}
\label{alg:weighted_swap}
Let $\A^{(0)}$ be the observed matrix\; 
\For{$k$ in 1:N}{
    Set $\A^{(k)} = \A^{(k-1)}$\;
    Sample two elements $(i, j)$ and $(i', j')$ uniformly at random from the non-zero elements of $\A^{(k)}$\;
    \If{$\A^{(k)}_{i'j} = 1$ or $\A^{(k)}_{ij'} = 1$}{
        next $k$\ (no checkerboard);
    }
    Sample $s \sim \text{Bernoulli}(p_{ij;i'j'})$\;
    \If{$s = 1$}{
        set $\A^{(k)}_{ij} = \A^{(k)}_{i'j'} = 0$ and $\A^{(k)}_{i'j} = \A^{(k)}_{ij'} = 1$\ (perform swap);
    }
}
\end{algorithm}

If all weights are positive, then we have the following result:

\begin{theorem}
Given binary matrix $\A$ and weight matrix $\W$ with $w_{ij}>0$, then Algorithm~\ref{alg:weighted_swap} generates a Markov chain with stationary distribution given by $\eqref{eq:model}$.
\label{thm:weighted_swap}
\end{theorem}

\noindent
Proof of Theorem~\ref{thm:weighted_swap} is given in Appendix~\ref{sec:swap_proof}.
If weights are allowed to equal zero, the Markov chains may mix more slowly and even become reducible as we demonstrate in the next section.

\subsection{Structural Zeros}
\label{subsec:struct_zeros}

As mentioned, structural zeros have received little attention since originally introduced by Connor \& Simberloff~\cite{connor_assembly_1979}.
We show here that structural zeros can be problematic for swapping algorithms, first addressing sampling efficiency then correctness.

Structural zeros can result in some checkerboard swaps having zero probability, which removes transitions from the Markov chain.
This can increase the number of swaps necessary to transition between two states, in turn reducing the sampling efficiency, and is reflected in a larger diameter of the Markov state space.
Here we show that Brualdi's~\cite{brualdi_matrices_1980} upper bound on the diameter of the Markov space no longer holds with structural zeros.

Given matrices $\A, \B \in \arc$ the upper bound on the maximum number of swaps required to transition from one matrix to the other is given by $d_H(\A,\B)/2 - k$, where $d_H(\A,\B)$ is the Hamming distance between $\A$ and $\B$, and $k$ is the largest number of non-overlapping ``elementary circuits'' in $\A-\B$, which we briefly explain here.
Consider $\A - \B$ as a weighted adjacency matrix of a bipartite graph, where the rows and columns form  separate partitions of a vertex set. 
Suppose an entry of $+1$ indicates an edge from the row vertex to the column vertex, and an element of  $-1$ indicates an edge from column vertex to the row vertex. 
An elementary circuit is then defined as a cycle of edges in this graph.

To show how the bound can be violated, consider the following matrices, both with unit row and column sums where structural zeros are represented with grey fill:

\vspace{4pt}
\begin{center}
$\A = $
\begin{tabular}{|c|c|c|c|c|}\hline
    1 & 0 & 0\sz & 0\sz & 0\sz \\\hline
    0 & 1 & 0 & 0\sz &  0\sz \\\hline
    0 & 0\sz & 1 & 0 & 0\sz \\\hline
    0 & 0\sz & 0\sz & 1 & 0\\\hline
    0 & 0 & 0 & 0\sz & 1 \\\hline
\end{tabular}
\quad \quad
$\B = $
\begin{tabular}{|c|c|c|c|c|}\hline
    1 & 0 & 0\sz & 0\sz & 0\sz \\\hline
    0 & 1 & 0 & 0\sz &  0\sz \\\hline
    0 & 0\sz & 0 & 1 & 0\sz \\\hline
    0 & 0\sz & 0\sz & 0 & 1\\\hline
    0 & 0 & 1 & 0\sz & 0 \\\hline
\end{tabular}
\end{center}
\vspace{10pt}

\noindent
The Hamming distance between them is $d_H(\A, \B) = 6$ and there is a single elementary circuit, giving a bound of $\frac{6}{2}-1 = 2$.
However, seven swaps are required to transition from $\A$ to $\B$, which can be described as a sequence of column pairs, since there is only one 1 per column: $\left( (1, 2), (1, 3), (1, 4), (1, 5), (1, 2), (2, 3) \right)$.

Additionally, the following example from Rao et al~\cite{rao_markov_1996} illustrates how structural zeros can render it impossible to transition between two matrices via checkerboard swaps:

\vspace{4pt}
\begin{center}
$\A = $
\begin{tabular}{|c|c|c|}\hline
    0\sz & 1 & 0 \\\hline
    0 & 0\sz & 1 \\\hline
    1 & 0 & 0\sz \\\hline
\end{tabular}
\quad \quad
$\B = $
\begin{tabular}{|c|c|c|}\hline
    0\sz & 0 & 1 \\\hline
    1 & 0\sz & 0 \\\hline
    0 & 1 & 0\sz \\\hline
\end{tabular}.
\end{center}
\vspace{10pt}

\noindent
Because no sequence of swaps transforms $\A$ into $\B$, the Markov chain described in Algorithm~\ref{alg:weighted_swap} is reducible and hence there is no unique stationary distribution.

In many cases, the presence of a few structural zeros will not result in a reducible chain, but in the general case structural zeros can be fatal to checkerboard swap based MCMC sampling algorithms.
However, let us consider a special case where they are not problematic, specifically where structural zeros satisfy the following:

\begin{definition}
The structural zeros of a matrix $\A$ are \textit{monotonic} if there exists a permutation of the rows and columns of $\A$, such that $w_{ij} = 0$ implies $w_{i'j} = w_{ij'} = 0$ for $i'<i,~j'<j$.
\end{definition}

This situation can arise when the rows and columns reflect different moments in time.
For example, let $\A$ represent a citation network among academic papers, where a 1 indicates that the row article cites the column article.  Suppose the rows and columns are arranged chronologically with increasing row/column index.
Clearly, articles cannot cite into the future, so we prohibit some 1's using structural zeros.
In this example, the upper triangle (including the diagonal) of $\A$ are structural zeros.  Furthermore these structural zeros are monotonic as $\A$ will satisfy the definition if the order of the columns of $\A$ are reversed.

When structural zeros are monotonic, Algorithm~\ref{alg:weighted_swap} produces a valid sampler:
\begin{theorem}
Given binary matrix $\A$ and weight matrix $\W$ where any structural zeros are monotonic, then Algorithm~\ref{alg:weighted_swap} generates a Markov chain with stationary distribution given by $\eqref{eq:model}$.
\label{thm:weighted_swap_sz}
\end{theorem}
The only modification of the proof for Theorem~\ref{thm:weighted_swap} needed to prove Theorem~\ref{thm:weighted_swap_sz} is to show irreducibility of the chain given structural zeros (see Appendix~\ref{sec:monotonic_proof}).

\subsection{Weighted Curveball Algorithm}

Strona et al.~\cite{strona_fast_2014} proposed a faster version of checkerboard swapping called the curveball, which imagines two rows trading their 1's like baseball cards. 
Two rows are randomly selected, and the 1's in their symmetric difference become candidates for the trade, where several 1's may move between rows in a trade.
In the uniform setting, the proposed trade is always performed.
In the non-uniform setting, we modify the algorithm in the same way as for weighted checkerboard swapping, and introduce a probabilistic trade, with probability dependent on the weights of the elements involved in the trade.
The entire procedure is presented in Algorithm~\ref{alg:weighted_curveball}.

\begin{algorithm}
\caption{Sampling Via Weighted Curveball}
\label{alg:weighted_curveball}
Let $\A^{(0)}$ be the observed matrix\; 
\For{$k$ in 1:N}{
    Set $\A^{(k)} = \A^{(k-1)}$\;
    Sample two rows $i$ and $i'$ (such that $i \neq i'$) uniformly at random from the non-zero elements of $\A^{(k-1)}$ \;
    Let $\A_{i \setminus i'} = \{j : \A_{ij}=1, \A_{i'j}=0, w_{i'j}>0 \}$ \;
    Let $\A_{i' \setminus i} = \{j : \A_{i'j}=1, \A_{ij}=0, w_{ij}>0 \}$ \;
    \If{$|\A_{i' \setminus i}| = 0$ or $|\A_{i \setminus i'}| = 0$}{
        next $k$\ (no trade);
    }
    \
    Let $C$  be the ordered list resulting from concatenating $\A_{i \setminus i'}$ and $\A_{i' \setminus i} $ \;
    Permute the elements of $C$ uniformly at random \;
    Let $\B_{i\setminus i'}$ be the first $|\A_{i\setminus i'}|$ elements of $C$, and let $\B_{i'\setminus i}$ be the last $|\A_{i'\setminus i}|$ elements of $C$ \;
    Let $\textbf{j}_{i} = \{j : j \notin \A_{i\setminus i'}, j \in \B_{i\setminus i'}$ \} (0 to 1 in row $i$) \;
    Let $\textbf{j}_{i'} = \{j : j \notin \A_{i'\setminus i}, j \in \B_{i'\setminus i}$ \} (0 to 1 in row $i'$) \;
    Compute $ p_{ii'} = \frac{\displaystyle \prod_{j \in \textbf{j}_{i}} w_{ij} \prod_{j \in \textbf{j}_{i'}} w_{i'j}}{\displaystyle \prod_{j \in \textbf{j}_{i}} w_{ij} \prod_{j \in \textbf{j}_{i'}} w_{i'j} + \prod_{j \in \textbf{j}_{i}} w_{i'j} \prod_{j \in \textbf{j}_{i'}} w_{ij}}$ \;
    Sample $s \sim \text{Bernoulli}(p_{ii'})$\;
    \If{$s = 1$}{
        Set $\A^{(k)}_{ij} = 1$ and $\A^{(k)}_{i'j} = 0$ for $j \in \textbf{j}$ \;
        Set $\A^{(k)}_{ij} = 0$ and $\A^{(k)}_{i'j} = 1$ for $j \in \textbf{j'}$ \;
    }
}
\end{algorithm}

Carstens~\cite{carstens_proof_2015} proves that the unweighted curveball algorithm can be used to sample uniformly, and we contribute the following theorem for the weighted case:
\begin{theorem}
Given binary matrix $\A$ and weight matrix $\W$ where any structural zeros are monotonic, then Algorithm~\ref{alg:weighted_curveball} generates a Markov chain with stationary distribution given by $\eqref{eq:model}$.
\label{thm:weighted_curveball}
\end{theorem}
In Appendix~\ref{sec:curveball_proof}, we provide a proof of this theorem when there are no structural zeros, and in  Appendix~\ref{sec:monotonic_proof} this is extended to the case of monotonic structural zeros.
\textsf{R} code implementing Algorithms~\ref{alg:weighted_swap} and~\ref{alg:weighted_curveball} is provided on Github at \emph{(https://github.com/fouticus/nusfimbim)}.

\section{Simulations}

We performed simulations to investigate the behavior of the weighted swap and weighted curveball algorithms.
In the first simulation, we verify the weighted checkerboard swap algorithm samples correctly from a non-uniform distribution for a small example.
In the second simulation, we compare the mixing performance of both algorithms for different weighting and structural zero schemes.
Finally, we investigate the effect of different weighting schemes on the resultant sampling distribution of a summary statistic.

For the second and third simulations, we will make use of a global statistic termed \emph{diagonal divergence}, which quantifies how far the 1’s of a square $(n \times n)$ matrix $\A$ are from its diagonal:
\begin{equation}
    T(\A) = \frac{1}{|\A|n}\sum_{ij} |i-j| \mathbb{I}(\A_{ij} = 1)
\end{equation}
where $|\A|$ is the number of 1's in $\A$ and $\mathbb{I}(\cdot)$ is the indicator function.

\subsection{Small Example}
We first consider a small example, such that every Markov state can be enumerated and exact probabilities can be computed. 
Consider the following binary matrix $\A$, and weight matrix $\W$:

\begin{align}
\A = 
\begin{tabular}{|c|c|c|}\hline
    1 & 0 & 0 \\\hline
    0 & 1 & 0 \\\hline
    0 & 0 & 1 \\\hline
\end{tabular}
\quad \quad
\W = 
\begin{tabular}{|c|c|c|}\hline
    1 & 2 & 1 \\\hline
    2 & 1 & 2 \\\hline
    1 & 2 & 1 \\\hline
\end{tabular}.\label{eq:toy_A_W}\\
\notag
\end{align}
Including $\A$, there are six possible states with row and column sums equal to 1 (see Figure~\ref{fig:toy_states}).
Under a uniform distribution, all matrices would have probability $\frac{1}{6}$. $\W$ induces a non-uniform probability distribution among the matrices (see Table~\ref{tab:toy_results}), where matrices $\A$ and $\D$ have lower probability because their non-zero elements correspond to 1's in the weight matrix.
We compare the true distribution to the distribution of a sample produced by weighted swapping.
To sample, we used a burn-in  of $10,000$ swaps before retaining every $10,000$th matrix, to generate a sample of $1,000$ matrices.
\begin{figure}
    \begin{center}
    $\A = $
    \begin{tabular}{|c|c|c|}\hline
        1 & 0 & 0 \\\hline
        0 & 1 & 0 \\\hline
        0 & 0 & 1 \\\hline
    \end{tabular}
    \quad \quad
    $\B = $
    \begin{tabular}{|c|c|c|}\hline
        0 & 1 & 0 \\\hline
        1 & 0 & 0 \\\hline
        0 & 0 & 1 \\\hline
    \end{tabular}
    \quad \quad
    $\CC = $
    \begin{tabular}{|c|c|c|}\hline
        1 & 0 & 0 \\\hline
        0 & 0 & 1 \\\hline
        0 & 1 & 0 \\\hline
    \end{tabular} \\
    \vspace{4pt}
    $\D = $
    \begin{tabular}{|c|c|c|}\hline
        0 & 0 & 1 \\\hline
        0 & 1 & 0 \\\hline
        1 & 0 & 0 \\\hline
    \end{tabular}
    \quad \quad
    $\E = $
    \begin{tabular}{|c|c|c|}\hline
        0 & 1 & 0 \\\hline
        0 & 0 & 1 \\\hline
        1 & 0 & 0 \\\hline
    \end{tabular}
    \quad \quad
    $\F = $
    \begin{tabular}{|c|c|c|}\hline
        0 & 0 & 1 \\\hline
        1 & 0 & 0 \\\hline
        0 & 1 & 0 \\\hline
    \end{tabular}
    \end{center}
    \caption{All $3 \times 3$ matrices with row and column sums equal to 1.}
    \label{fig:toy_states}
\end{figure}
Table~\ref{tab:toy_results} shows agreement between the true and empirical probability distributions. 
The KL-Divergence between the empirical and true distributions is $9.0\times 10^{-4}$.

\begin{table}[]
    \centering
    \begin{tabular}{c c c}\hline
        State &  Probability & \makecell{Empirical \\ Probability} \\ \hline
        $\A$  & 0.056 & 0.051 (+/- 0.007) \\\hline
        $\B$  & 0.222 & 0.237 (+/- 0.013) \\\hline
        $\CC$ & 0.222 & 0.217 (+/- 0.013) \\\hline
        $\D$  & 0.056 & 0.058 (+/- 0.008) \\\hline
        $\E$  & 0.222 & 0.222 (+/- 0.013) \\\hline
        $\F$  & 0.222 & 0.215 (+/- 0.014) \\\hline
    \end{tabular}
    \caption{Theoretical vs. empirical probabilities for matrices in Figure~\ref{fig:toy_states} with binary matrix and weight matrix as defined in \eqref{eq:toy_A_W}. Empirical probabilities also show Tukey-Hanning~\cite{flegal_batch_2010} MCMC standard errors. }
    \label{tab:toy_results}
    %The agreement of distributions reflects correct weighted sampling.
\end{table}

\subsection{Mixing Performance}
\label{subsec:mixing}
In MCMC sampling, efficient mixing is desirable in order to maximize inferential capability and minimize computation time. 
Due to autocorrelation in sequential observations, the precision of a statistic from MCMC sampling is less than it would be for independent, identically distributed (iid) sampling. 
The difference in precision can be measured by effective sample size (ESS), which gives the number of iid samples required to match the precision of the MCMC sample~\cite{hoff_first_2009}. 
We simulated matrices using weighted checkerboard swapping and weighted curveball under different weight schemes and structural zero configurations in order to assess their impact on mixing, as measured by ESS.
ESS was computed for the diagonal divergence statistic using the \textit{coda} package~\cite{plummer_coda_2006}.

To assess the impact of weights, we constructed $20 \times 20$  weight matrices by sampling each element independently from one of the distributions: Exponential(1), Uniform(0,1), and Uniform(0.5, 1), and also a weight matrix of all 1’s, corresponding to uniform sampling. 
After constructing the weight matrices, we introduced structural zeros with one of two methods. 
The first method was to randomly set elements equal to 0 with fixed probability $p$, where $p$ was either 0.1, 0.25, or 0.5. 
The resulting matrices will likely correspond with irreducible Markov chains, given the relatively small probabilities.
The second method was to designate a lower triangular region of the weight matrix as all zeros, so that they would be monotonic. 
Triangles covered either 10\%, 25\%, or 50\% of the weight matrix. We generated a $20 \times 20$ binary starting matrix by setting each element to 1 independently with probability 0.25, resulting in a density of 23.5\% before enforcing structural zeros. %, since elements can’t be 1 if the corresponding weight is 0.

\begin{figure}
    \centering
    \includegraphics[width=0.95\linewidth]{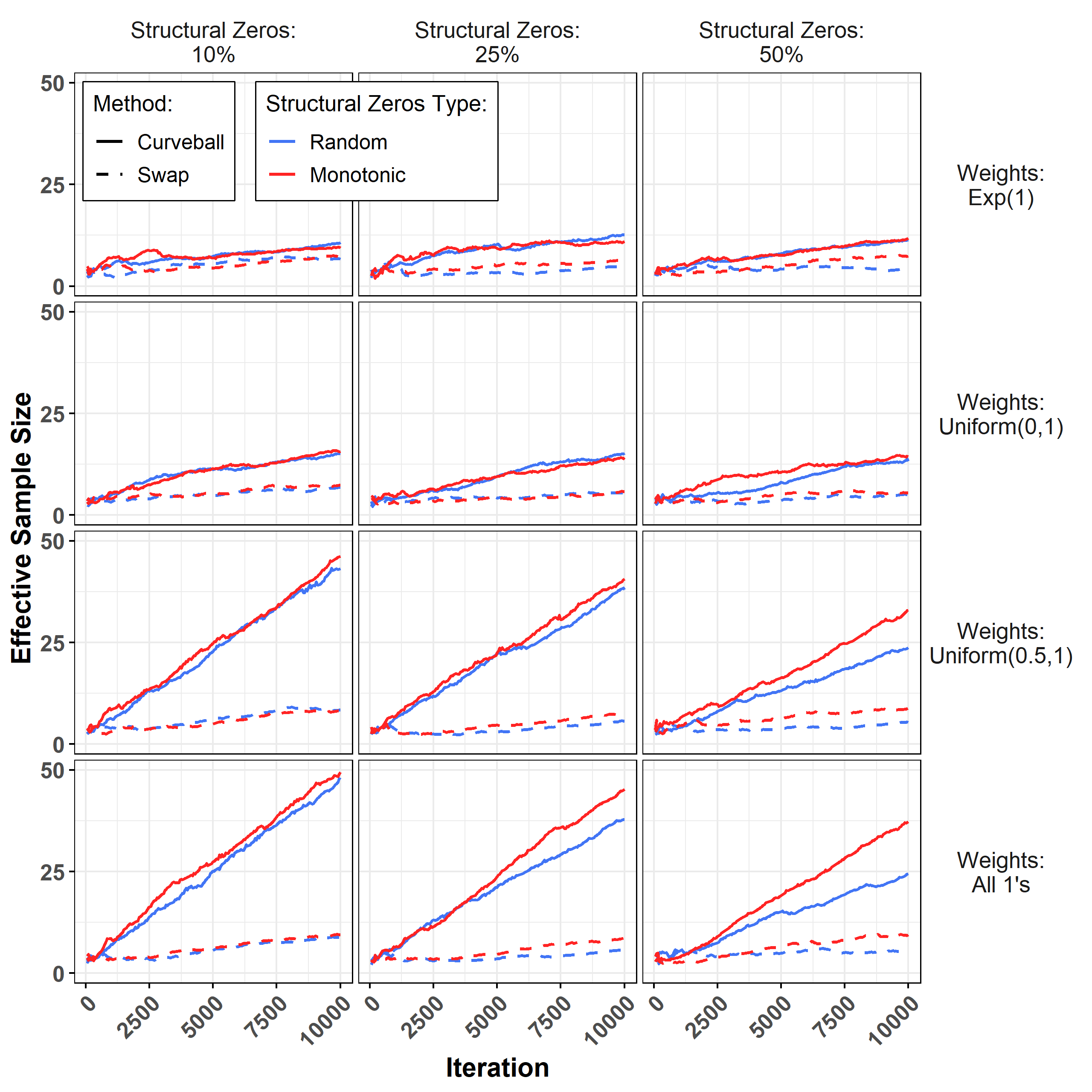}
    \caption{Average effective sample size for the simulations described in Section~\ref{subsec:mixing}, for varying weighting schemes and percentage of structural zeros, where structural zeros are either randomly placed (blue), or arranged monotonically (red). Each line represents the average ESS from 10 separate chains using the same starting matrix.}
    \label{fig:mixing_facet}
\end{figure}

\begin{figure}
    \centering
    \includegraphics[width=0.95\linewidth]{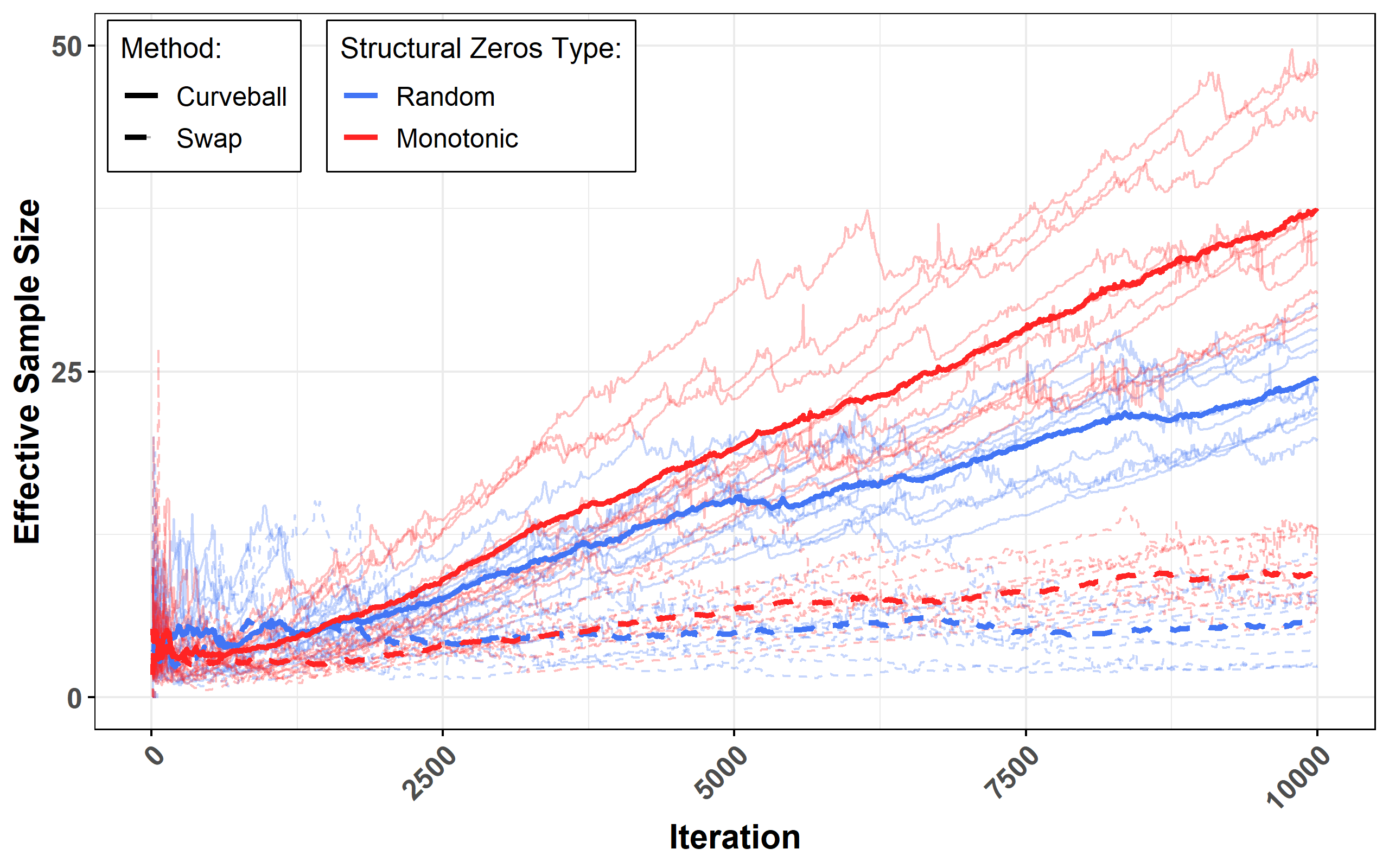}
    \caption{Detailed view of effective sample size for the simulations described in Section~\ref{subsec:mixing}, using a weight matrix of all 1's and with 50\% structural zeros. Average ESS is shown in bold, with each iteration shown partially transparent.}
    \label{fig:mixing_iterations}
\end{figure}

For each combination of weights and structural zeros, we ran each sampling algorithm for 10,000 iterations (with no burn-in or thinning). %, since the purpose is to observe mixing performance. 
Figure~\ref{fig:mixing_facet} shows ESS for the different conditions, where each line is the average over 10 individual chains with different random seeds. 
For illustration, Figure~\ref{fig:mixing_iterations} shows ESS for all chains for the condition in the bottom-right panel of Figure~\ref{fig:mixing_facet}.

Comparing the weighted swap algorithm to the weighted curveball algorithm, we see that the curveball algorithm tends to mix more rapidly. 
This reflects the fact that weighted curveball can perform multiple trades simultaneously, whereas weighted swap only swaps two 1’s at a time, increasing the autocorrelation between sequential observations. 
This difference is less pronounced for the Exponential(1) and Uniform(0,1) weighting schemes compared to the Uniform(0.5, 1) and ``All 1’s’’ weighting schemes. 
This is because the former weight schemes contain weights near 0, making 1 in the corresponding position of $\A$ possible but unlikely. Since curveball attempts to make several trades simultaneously, this increases the chances of attempting to trade a 1 into an unlikely position. 
This makes the trade itself unlikely, so fewer trades take place overall, removing the advantage that weighted curveball has over weighted swap. 
The latter two weight schemes do not produce weights near 0 (ignoring structural zeros), so more curveball trades are successful, restoring the ESS gap between the algorithms. 
One instance where weighted swap may have an advantage over weighted curveball is in sparse matrices with weights near 0. The sparsity would limit the ability of weighted curveball to perform multiple trades, and the low weights would favor the ``smaller steps’’ taken by weighted swap. 

Comparing random structural zeros to monotonic structural zeros, we see little difference for the top two weighting schemes. 
As mentioned before, these two schemes already tend to produce matrices with weights near 0, so setting some elements equal to zero does not change the weight matrices much, whether done randomly or monotonically. 
However, the bottom two weight schemes produce weights away from 0, so setting some as structural zeros is a bigger change. In these cases, mixing is better when structural zeros are monotonic compared to random, with the difference being more pronounced when there are more structural zeros. 
This is unsurprising given the discussion in Section~\ref{subsec:struct_zeros} showing that structural zeros may give rise to reducible chains, whereas for monotonic structural zeros irreducibility is preserved, and the space has bounded diameter (see Appendix C).

Finally, we acknowledge that ``well mixing’’ may be a questionable label for any of the scenarios shown, since ESS is relatively paltry compared to the number of iterations. In practice, this is ameliorated by careful algorithm implementation (which we do not discuss here) and aggressive thinning.

\subsection{Effect of Heterogeneous Weights}
\label{sec:effect_of_weights}
\begin{figure}
    \centering
    \begin{subfigure}{0.5\linewidth}
    \centering
        \includegraphics[width=0.5\linewidth]{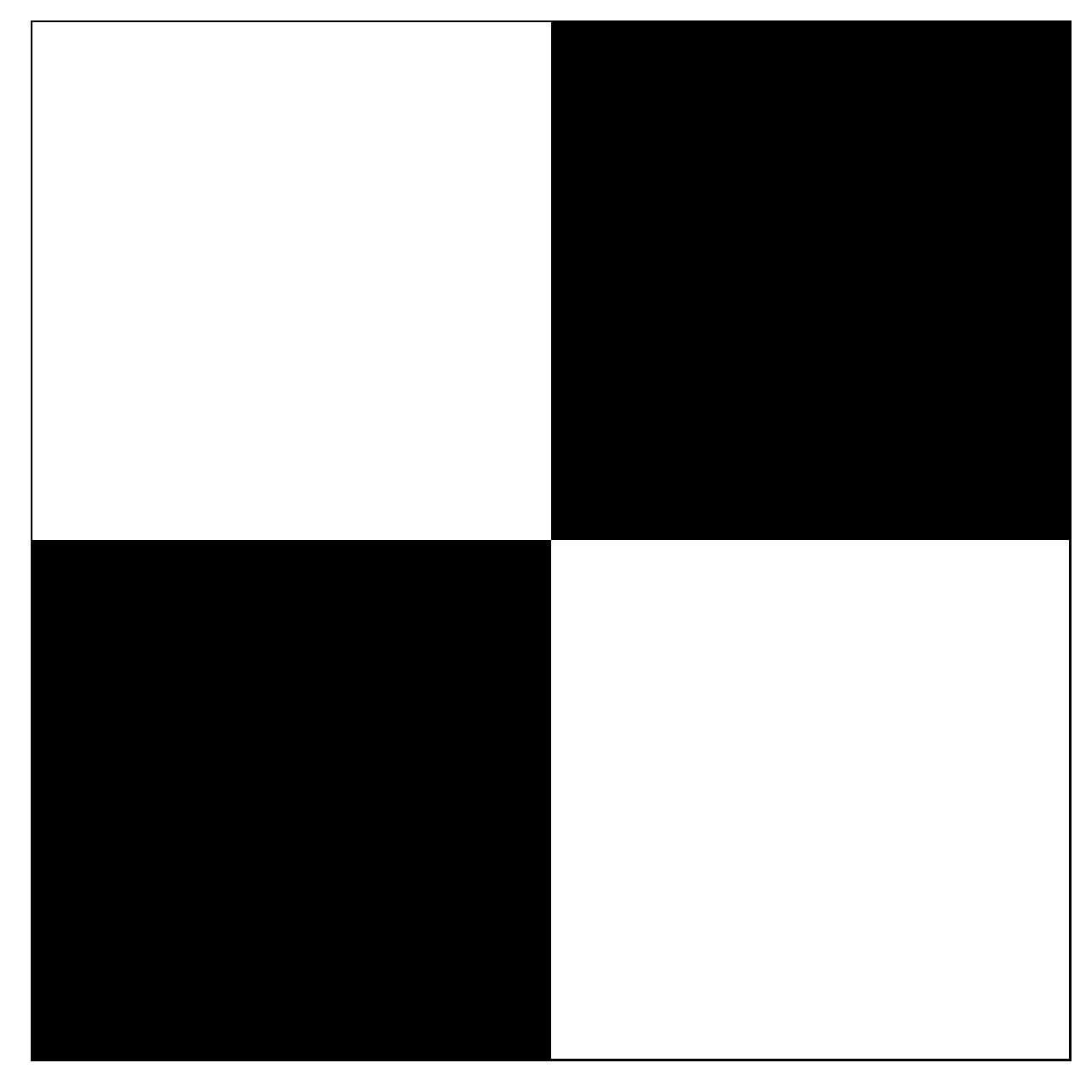}
        \caption{Initial matrix, $\A$}
        \label{fig:wp_A}
    \end{subfigure}%
    \begin{subfigure}{0.5\linewidth}
    \centering
        \includegraphics[width=0.5\linewidth]{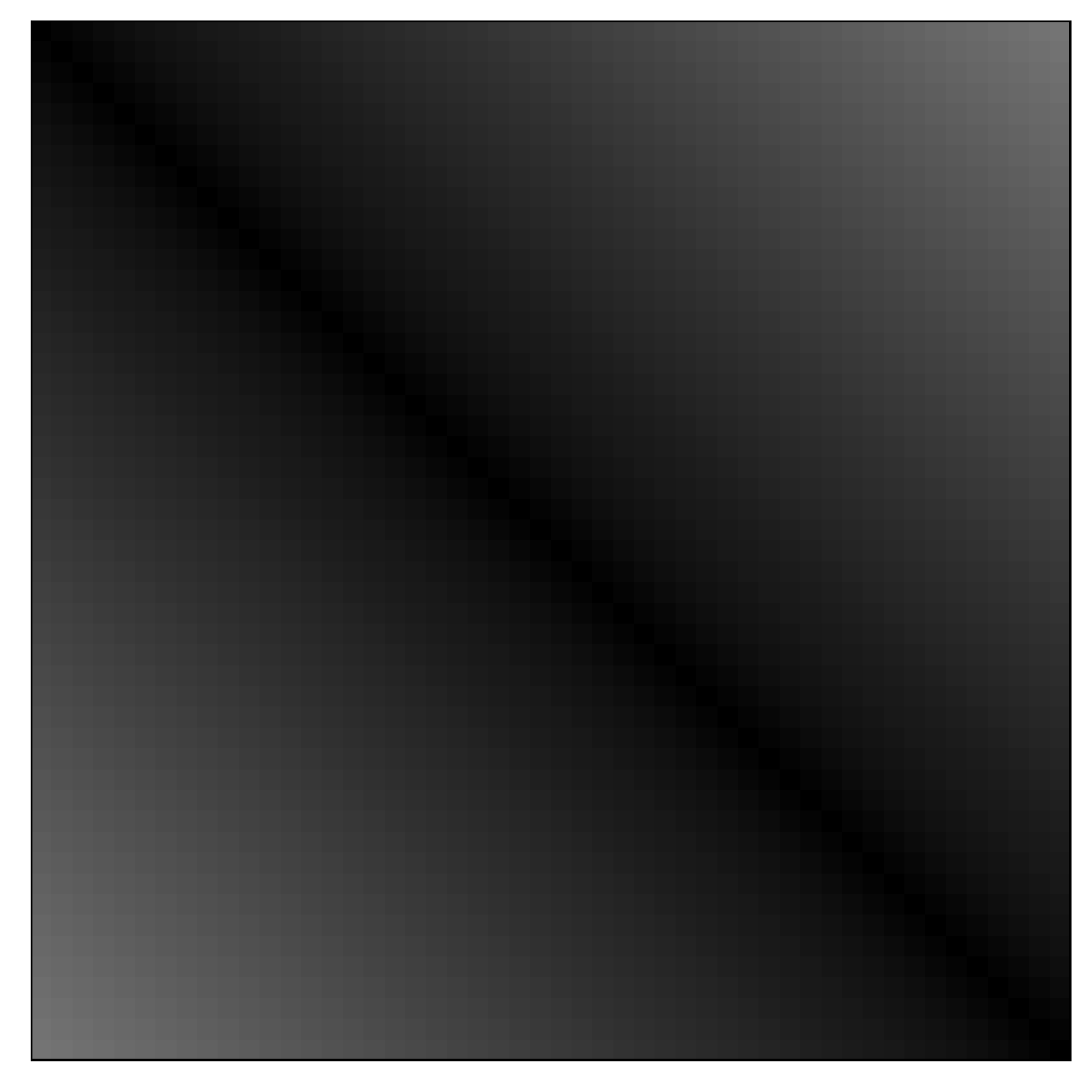}
        \caption{Base weight matrix, $\W$}
        \label{fig:wp_W}
    \end{subfigure}
    \caption{Initial binary matrix $\A$ and base weight matrix $\W$ for the simulation described in Section~\ref{sec:effect_of_weights}. Both use a shading scale where white denotes 0 and black denotes 1. The minimum value in $\W$ is 0.5, achieved in the top-right and bottom-left most elements. Note the unfavorable state of $\A$, since it contains 1's corresponding to lower weighted regions of $\W$.}
    \label{fig:wp}
\end{figure}

The matrix $\W$ adds flexibility to the sampling algorithms by allowing for non-uniform stationary distributions, however it is unclear how much influence $\W$ has over the resultant sample given the fixed row and column margins.
To investigate this issue, we compare the sampling distribution for a statistic under different ``strength'' weight matrices, which are constructed by taking the elements of a base weight matrix to a positive or negative power.
In this case, we let $\A$ have dimension $50 \times 50$, and define the base weight matrix as:
\begin{equation}
    \W_{ij} = \frac{100-|i-j|}{100},
\end{equation}
which has value $1$ on the diagonal and value near $0.5$ in the corners (see Figure~\ref{fig:wp}).
%We compare samples taken under various weight matrices which are elementwise powers of the base weight matrix. 
Sampling proceeded with a burn-in period of $5,000$ proposed swaps, before retaining every $1,000$th matrix, to generate a sample of $5,000$ retained matrices.

\begin{figure}
    \centering
    \includegraphics[width=0.99\linewidth]{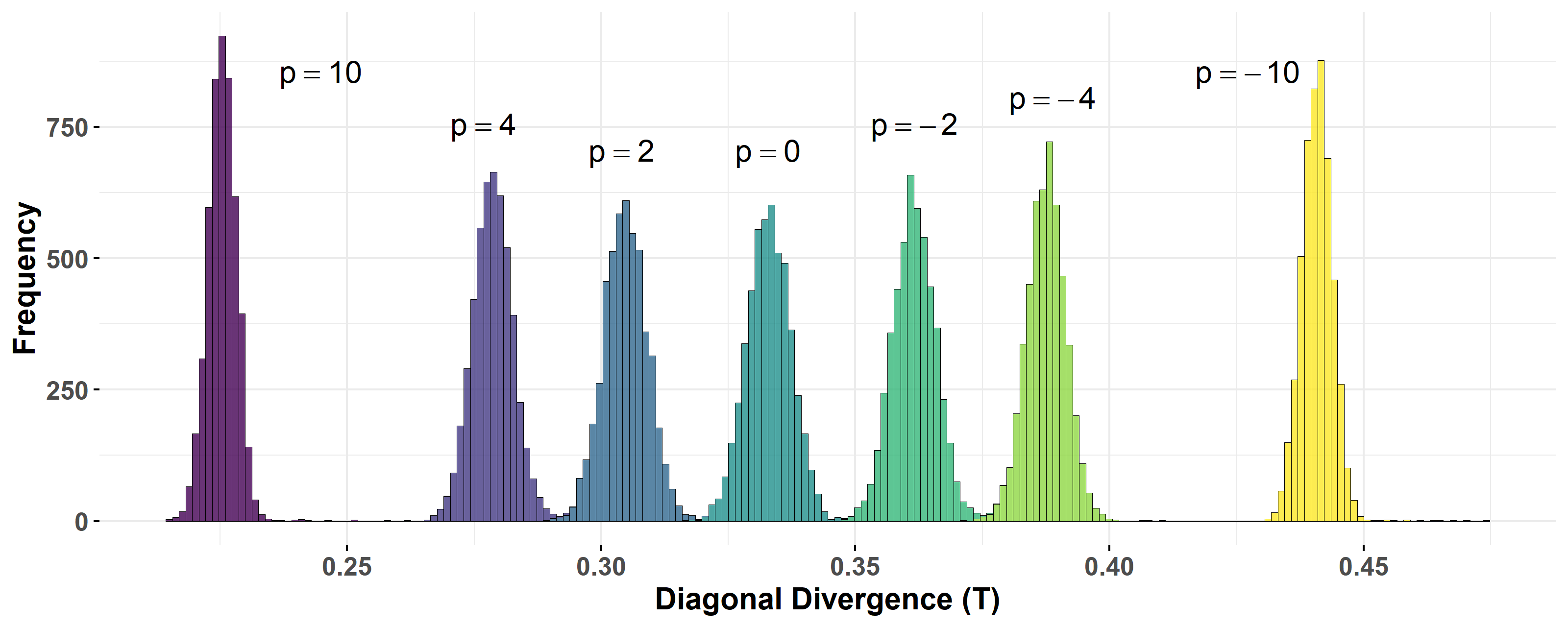}
    \caption{Sampling distribution of the diagonal divergence statistic, under different weighting matrices, defined by raising the base weight matrix to the power $p$.}
    \label{fig:weights}
\end{figure}

Figure~\ref{fig:weights} shows the approximate sampling distributions of the diagonal divergence $T$ for selected powers of the weight matrix. Since $\W$ favors matrices with 1's near the diagonal, larger powers of $\W$ favor lower values for $T$, and negative powers of $\W$ favor higher values for $T$. Larger values of $|p|$ emphasize having 1's close to the diagonal ($p$ positive) or packed in the corners ($p$ negative), where there are relatively few distinct matrices, and there is a corresponding decrease in the variance of the samples in those cases. This figure illustrates how vast the space $\arcp$ is as there is virtually no overlap in the sampling distributions under different weighting schemes.

\section{Example: New Hebrides Bird Species}
\label{sec:new_hebrides}
\begin{figure}
    \centering
    \begin{subfigure}{0.4\linewidth}
    \centering
        \includegraphics[width=\linewidth]{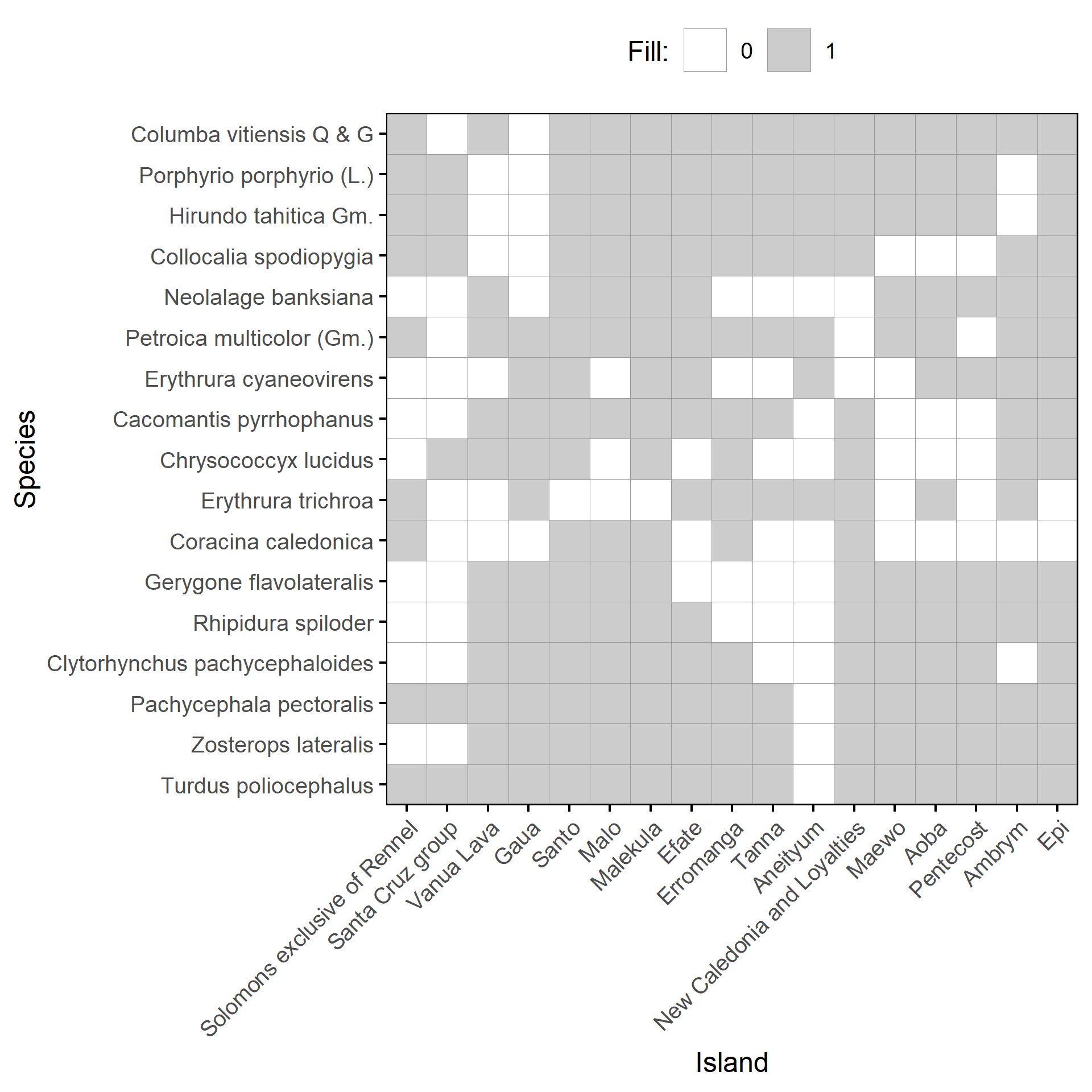}
        \caption{Species-Island occurrence, $\A$}
        \label{fig:nh_A}
    \end{subfigure}
    \begin{subfigure}{0.4\linewidth}
    \centering
        \includegraphics[width=\linewidth]{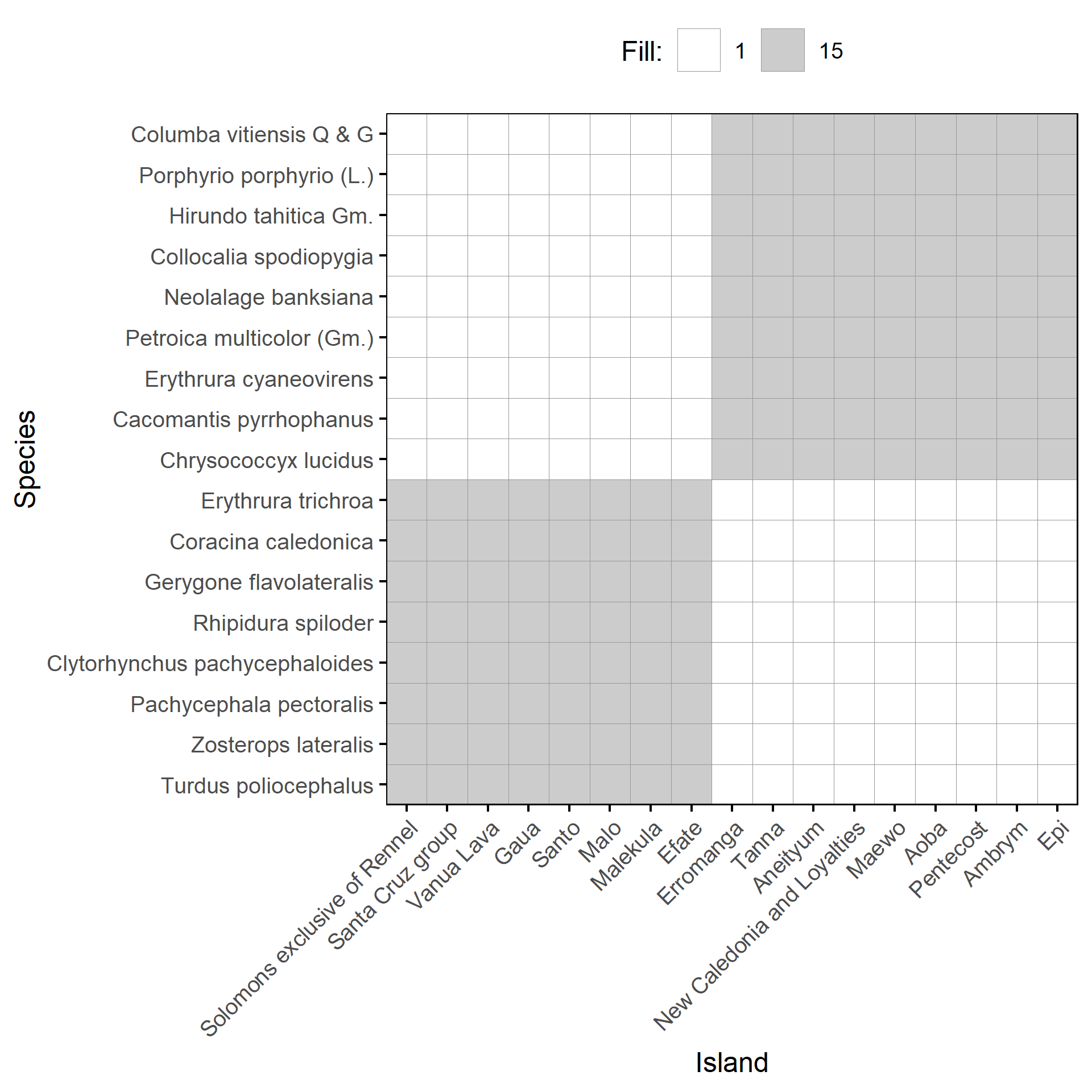}
        \caption{Weight matrix, $\W$}
        \label{fig:nh_W}
    \end{subfigure}
    \caption{Observed New Hebrides bird species combination $\A$~\cite{diamond_assembly_1975}, and non-uniform weight matrix $\W$ for the data analysis described in Section~\ref{sec:new_hebrides}. For $\A$, white denotes 0 and grey denotes 1. For $\W$, light grey denotes 1 and grey denotes 15. With higher values indicating species (rows) preference for certain island (columns) habitats.}
    \label{fig:nh}
\end{figure}

We revisit the original Vanuatu (formerly New Hebrides) bird species data provided by Diamond in 1975~\cite{diamond_assembly_1975}, to illustrate the impact weighting can have on the analysis of real data. 
Of interest is whether or not there exists competitive exclusion between different bird species, or groups of species, on these islands.
That is, do certain species or groups exclude one another from islands due to competition over resources?
The data are represented in a matrix $\A$, where rows are species, and columns are islands (see Figure~\ref{fig:nh_A}).
The elements of the matrix are 1 if the corresponding species is found on the corresponding island, and 0 otherwise.
If a checkerboard pattern exists between two species on two islands, this can indicate competitive exclusion~(see~\cite{roberts_island-sharing_1990}).
Accordingly, Stone \& Roberts~\cite{stone_checkerboard_1990} propose the \textit{C-score} to measure competitive exclusion by counting how many checkerboards exist between species pairs in the matrix:
\begin{equation}
C(\A) = \frac{2}{m(m-1)} \sum_{j=2}^m \sum_{i<j} (\Rs_i - S_{ij})(\Rs_j - S_{ij}).
\label{eq:c_score}
\end{equation}
Here, $S_{ij}$ is the number of shared islands between species $i$ and $j$, $\Rs_i$ is the number of islands inhabited by species $i$, and $m$ is the number of species.
A high C-score is interpreted as evidence of competitive exclusion.
We will analyze the data using a null model analysis twice, once with equal weights (as has been done historically), and once with heterogeneous-weights
Though we are using real data, this is an illustrative example only and we select a weight matrix to that end. 

We define a weight matrix $\W$ to encode the hypothesis that  island preference by species is higher for certain islands than others (see Figure~\ref{fig:nh_W}).
This could result, for example, from several environmental and biological factors.
We then proceed, as did Connor \& Simberloff~\cite{connor_assembly_1979}, to generate a sample of matrices under the null distribution, assuming equal weights (all 1's) and then under $\W$.
We generate $5,000$ samples from each weighting regime, after a burn-in of $1,000$, and thinning of $500$, using the weighted curveball algorithm.

\begin{figure}
    \centering
    \includegraphics[width=0.99\linewidth]{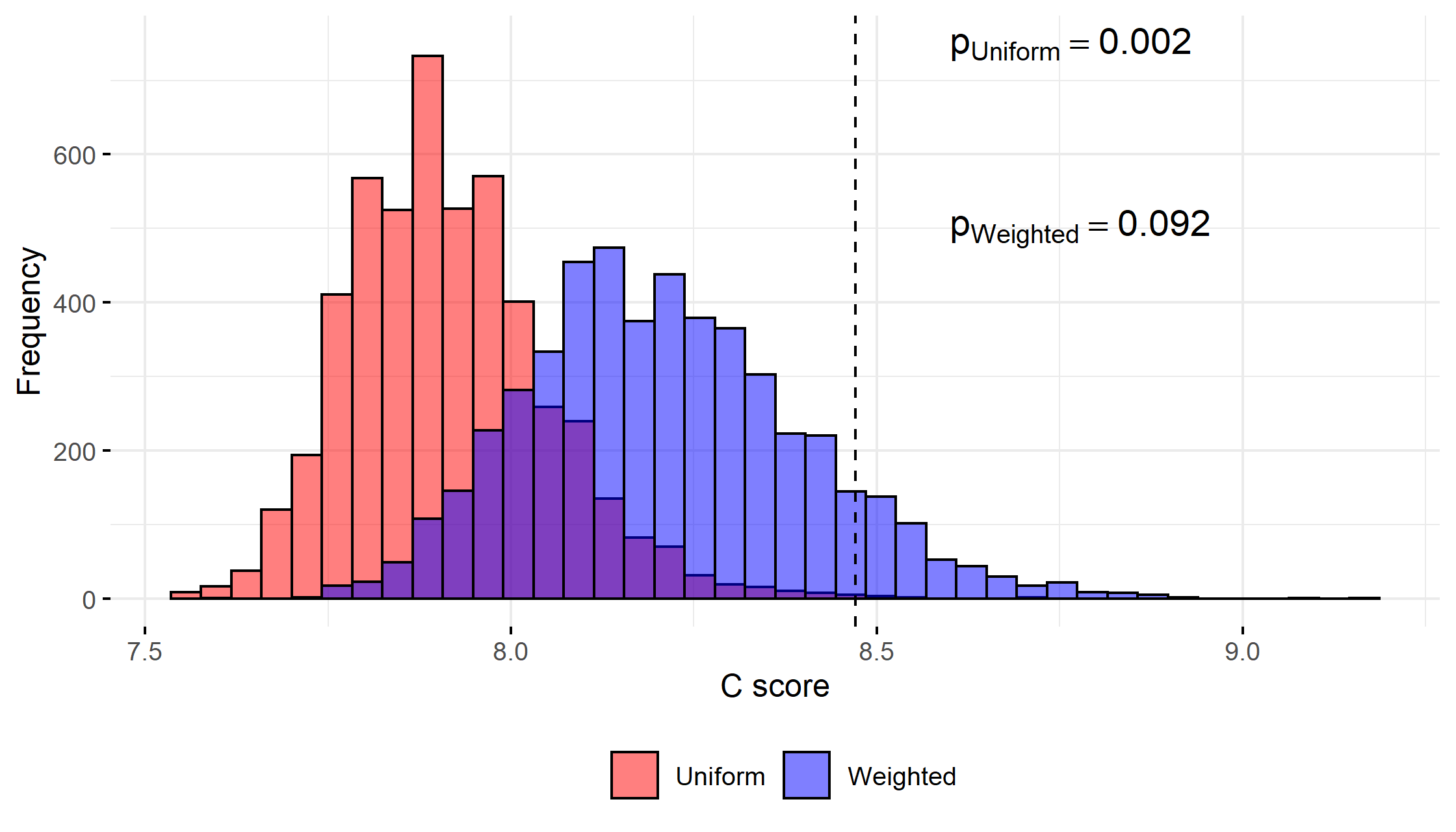}
    \caption{Sampling distribution of the C-score statistic, under uniform (red) and non-uniform (blue) weights. The empirical p-value under non-uniform weights is no longer significant, demonstrating how incorporating species habitat preference could impact whether the observed combination is relatively unlikely.}
    \label{fig:nh_hist}
\end{figure}

The resulting distributions of C-scores are shown in Figure~\ref{fig:nh_hist}, along with the empirical upper-tailed p-values of the observed matrix under each model.
We see significance at the $\alpha=0.05$ level under uniform sampling, but not after accounting for the weights.
The blocked structure of species-island preferences represented in $\W$ favors more checkerboard structures between species in the top half and bottom half, such that the observed matrix no longer appears so exceptional. 
The lesson here is that by incorporating more information about the bird species into our null model, the evidence in favor of competitive exclusion changes, and thus researchers should carefully consider the appropriate weight matrix when performing such analyses.

\section{Discussion}
Given the healthy debate surrounding the appropriate use of null matrix model analysis methods, we recommend that the practitioner weigh this approach against their particular situation and needs.
One advantage of this approach is that by constraining, equivalently conditioning on, the row and column sums of the matrix, we remove the need to explicitly model them.
However, without an estimation procedure for the weights, these values must be specified \emph{a priori}, and reasonable sensitivity analyses are warranted.
Fixing row and column sums is only one way to perform null model analysis, and as Gotelli stated~\cite{gotelli_null_2000}: ``There is great value in exploring the results of several null models that incorporate different degrees of randomness''.
These types of null models provide information about mechanism through the careful consideration of a summary statistic, but it often can be difficult to argue for the correspondence between mechanism and statistic directly.
Therefore, if mechanism is of importance, a direct modeling approach may be more appropriate.

We have introduced weighted versions of the checkerboard swap and curveball algorithms, which extend binary matrix sampling to non-uniform probability distributions.
Simulations demonstrate that the choice of weight matrix can impact the efficiency of sampling.
In fact, when weights are equal to 0, this can slow the mixing process of the MCMC chain, or even prohibit sampling from the desired distribution. 
However, monotonic structural zeros can arise in time-evolved networks and are better behaved.

Structural zeros have also received attention from Rao et al~\cite{rao_markov_1996} who noted that when structural zeros are on the matrix diagonal, the Markov space may be disconnected when limited to ``alternating rectangles''(checkerboard swaps).
Their solution was to also consider ``alternating hexagons'', which they prove connects the space.
Alternative swapping mechanisms like this can be incorporated into the non-uniform setting straightforwardly, by defining the swap probability as a ratio of the relevant weights.

Both the weighted swap and weighted curveball algorithms rely on local changes to perform an MCMC walk, which has implications for sampling efficiency (as seen in Figure~\ref{fig:mixing_facet}). 
An unavoidable consequence is that large numbers of swaps are required to generate near independent samples.
Of course, the computational cost of each swap is very small, and several chains can be run in parallel to produce larger samples.
In general, the weighted checkerboard swaps is preferred when the matrix is very sparse or very dense, and weighted curveball is preferred when the matrix is closer to 50\% dense (which will produce the most trade candidates). The choice of summary statistic also has implications for mixing.
A single ``global'' statistic, like the C-score, mixes faster than a local statistic, like, for example, the marginal probability for an element of the matrix.

Another possible area of application is directed graphs, which are naturally represented as square binary adjacency matrices.
Fixing marginal sums of these matrices is equivalent to fixing in-degree and out-degree for each vertex.
Unweighted checkerboard swapping has been used to sample uniformly for vertex-labelled networks with self-loops, also known as the Configuration Model~\cite{fosdick_configuring_2018}.
However, we know of no implementation of graph sampling where edges are weighted, as in this paper.

In this work, we do not address how to select appropriate weights, as this should be informed by the scientific setting and specific hypothesis being tested.
However, the following observation may help guide the construction of $\W$.
Consider binary matrices $\A$ and $\B$ which differ only by a checkerboard swap in rows $i$ and $i'$, and columns $j$ and $j'$:
\begin{equation}
    \frac{\P(\A)}{\P(\B)} = \frac{w_{ij} w_{i'j'}}{w_{i'j} w_{ij'}}.
\end{equation}
The ratio of weights communicates the relative probability of the differing elements of two matrices, conditioning on all other elements.
Therefore eliciting or estimating the relative propensity of matrix elements could inform the weights specification.  

Classic (unweighted) checkerboard swapping and curveball trading, while intuitive and simple to implement, lack the ability to incorporate heterogeneity of probabilities when sampling, rendering them useless in situations where such heterogeneity is known to exist.
With our contribution, these algorithms are now equipped to incorporate easy to interpret weights, and to consider more plausible null models, which provides a valuable new tool for matrix analysis.

\section{Acknowledgements}
This work was partially supported by grant IOS-1856229 from the National Science Foundation.

\bibliographystyle{ACM-Reference-Format}
\bibliography{references.bib}

\newpage   % blank page 10
\mbox{}

\appendix

\section{Proof of weighted checkerboard algorithm}
\label{sec:swap_proof}

Here we prove that the weighted checkerboard swap algorithm samples from $\P(\cdot)$ defined in \eqref{eq:model} when there are no structural zeros, by showing that the Markov chain implied by the algorithm is ergodic (irreducible and aperiodic) with stationary distribution $\P(\cdot)$~\cite{kannan_simple_1999,jerrum_markov_1996}.

\textbf{Irreducibility.}
Since $w_{ij}>0$ we have $0<p_{ij:i'j'}<1$, so irreducibility follows directly from the result of Brualdi~\cite{brualdi_matrices_1980}, which showed that a series of swaps could change any matrix $\A$ into any other matrix $\B$ when row and column sums are preserved.

\textbf{Aperiodicity.}
Since every checkerboard swap has probability less than 1, it is possible to remain in the same state after a step.
This is enough to show aperiodicity.

\textbf{Detailed Balance.}
Lastly, we show that $P(\cdot)$ is the stationary distribution by showing that it satisfies the detailed balance equation:
\begin{equation}
    \P(\A)~ p_{\A \rightarrow \B} = \P(\B) ~ p_{\B \rightarrow \A}.
    \label{eq:balance}
\end{equation}
for any matrices $\A, \B \in \arcp$, where $p_{\A \rightarrow \B}$ and $p_{\B \rightarrow \A}$ are transition probabilities.
If $\A$ and $\B$ do not differ by a checkerboard swap, then both transition probabilities are 0, and detailed balance holds trivially.
If $\A$ and $\B$ differ by a checkerboard swap, then let $u$ and $u'$ be its rows and let $v$ and $v'$ be its columns.
Also, let $k$ be the sum of the row counts, and define the set $\mathcal{S} = \{(u,v), (u',v), (u,v'), (u',v')\}$, so that $a_{ij} = b_{ij}$ for $(i,j) \notin \mathcal{S}$.
The transition probability $p_{\A \rightarrow \B}$ is the probability of first selecting the two 1's in the checkerboard, $\binom{k}{2}^{-1}$, times the probability of performing the swap $p_{uv;u'v'}$.
Starting from the left side of \eqref{eq:balance},

\begin{align}
    &\P(\A) p_{\A\rightarrow\B} = \\
    &= \left[ \frac{1}{\kappa} \prod_{ij} w_{ij}^{a_{ij}} \right] \left[ \binom{k}{2}^{-1} \frac{w_{u'v}w_{uv'}}{w_{uv}w_{u'v'} + w_{u'v}w_{uv'}} \right] \\
    &= \frac{1}{\kappa} \binom{k}{2}^{-1} \Bigg[\prod_{ij \notin \mathcal{S}} w_{ij}^{a_{ij}} \Bigg] w_{uv} w_{u'v'} \frac{w_{u'v}w_{uv'}}{w_{uv}w_{u'v'} + w_{u'v}w_{uv'}} \\
    &= \frac{1}{\kappa}\binom{k}{2}^{-1} \Bigg[\prod_{ij \notin \mathcal{S}} w_{ij}^{b_{ij}} \Bigg] w_{u'v}w_{uv'} \frac{w_{uv} w_{u'v'}}{w_{uv}w_{u'v'} + w_{u'v}w_{uv'}} \\
    &= \left[\frac{1}{\kappa} \prod_{ij} w_{ij}^{b_{ij}} \right] \left[ \binom{k}{2}^{-1} \frac{w_{u'v}w_{uv'}}{w_{uv}w_{u'v'} + w_{u'v}w_{uv'}} \right] \\
    &= \P(\B) p_{\B\rightarrow\A},
\end{align}
so detailed balance holds.

\section{Proof of weighted curveball algorithm}
\label{sec:curveball_proof}

Here we prove that the weighted curveball algorithm samples from $\P(\cdot)$ defined in \eqref{eq:model} when there are no structural zeros.

\textbf{Irreducibility.}
Any checkerboard swap can be viewed as a curveball trade, so the set of all weighted curveball trades contains the set of all weighted checkerboard swaps.
Therefore any state transition in the checkerboard swap Markov chain is also a state transition in the curveball trade chain, hence irreducibility holds for curveball as well.

\textbf{Aperiodicity.}
Again, since $w_{ij}>0$, then $0<p_{ii'}<1$ for any proposed trade, so the probability of remaining in the same state is non-zero, and the chain is aperiodic.

\textbf{Detailed Balance.}
For detailed balance, consider matrices $\A$ and $\B$ which differ by a curveball trade in rows $u$ and $u'$.
The curveball algorithm involves permuting the elements of the trade candidates in $\A_{u\setminus u'}$ and $\A_{u'\setminus u}$ to create a proposed trade, captured in $(\B_{u\setminus u'}, \B_{u'\setminus u})$.
The probability of obtaining $(\B_{u\setminus u'}, \B_{u'\setminus u})$ is
\begin{align}
p_c &= |A_{u\setminus u'}|!|A_{u' \setminus u}|!/\left( |A_{u \setminus u'}| + |A_{u' \setminus u}| \right)! \\ 
&= |B_{u\setminus u'}|!|B_{u' \setminus u}|!/\left( |B_{u \setminus u'}| + |B_{u' \setminus u}| \right)!,
\end{align}
where $|\cdot|$ is the cardinality operator.
Note that $p_c$ is the same for all such partitions, that is, all partitions are equally likely.
Let $\mathcal{S} = \{ij: i \in \{u, u'\}, j \in \{\textbf{j}_i \cup \textbf{j}_{i'}\}\}$, and for a generic index $t$ and index set $\textbf{s}$, define $w_{t\textbf{s}} = \prod\limits_{j \in \textbf{s}} w_{tj}$.
Again starting from the left side of \eqref{eq:balance}:
\begin{align}
    &\P(\A) p_{\A\rightarrow\B} = \P(\A)p_c p_{uu'} \\
    &= \frac{p_c}{\kappa} \prod_{ij} w_{ij}^{a_{ij}} \left[ \frac{w_{u\textbf{j}_i}  w_{u'\textbf{j}_{i'}} }{w_{u\textbf{j}_i} w_{u'\textbf{j}_{i'}} + w_{u'\textbf{j}_i} w_{u\textbf{j}_{i'}}}\right] \\
    &= \frac{p_c}{\kappa}  \left[ \prod_{ij \notin \mathcal{S}} w_{ij}^{a_{ij}} \right]  w_{u'\textbf{j}_i} w_{u\textbf{j}_{i'}} \left[ \frac{ w_{u\textbf{j}_i} w_{u'\textbf{j}_{i'}} }{w_{u\textbf{j}_i} w_{u'\textbf{j}_{i'}} + w_{u'\textbf{j}_i} w_{u\textbf{j}_{i'}}} \right] \\
    &= \frac{p_c}{\kappa}  \left[ \prod_{ij \notin \mathcal{S}} w_{ij}^{b_{ij}} \right]  w_{u\textbf{j}_i} w_{u'\textbf{j}_{i'}} \left[ \frac{ w_{u'\textbf{j}_i} w_{u\textbf{j}_{i'}} }{w_{u\textbf{j}_i} w_{u'\textbf{j}_{i'}} + w_{u'\textbf{j}_i} w_{u\textbf{j}_{i'}}} \right] \\
    &= \frac{p_c}{\kappa} \prod_{ij} w_{ij}^{b_{ij}} \left[ \frac{w_{u'\textbf{j}_i}  w_{u\textbf{j}_{i'}} }{w_{u\textbf{j}_i} w_{u'\textbf{j}_{i'}} + w_{u'\textbf{j}_i} w_{u\textbf{j}_{i'}}}\right] \\
    &= \P(\B) p_c p_{u'u} \\
    &= \P(\B) p_{\B\rightarrow\A}
\end{align}
\noindent
Hence the detailed balance equation holds, where the stationary distribution is again given by \eqref{eq:model}.

\section{Irreducibility With Monotonic Structural Zeros}
\label{sec:monotonic_proof}

\begin{figure}
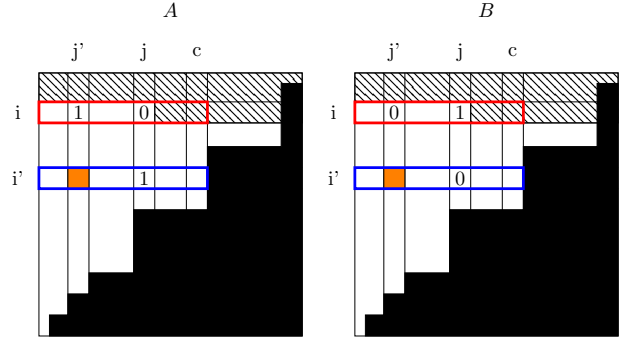

    \newcommand{\w}{5}
    \newcommand{\hw}{0.2}
    \newcommand{\jj}{\w*0.4}
    \newcommand{\jjj}{\w*0.15}
    \newcommand{\jc}{\w*0.6}
    \newcommand{\ii}{\w*0.15}
    \newcommand{\iii}{\w*0.4}
    \begin{tikscale}
        \coordinate (A) at (-\w-0.5, \w);
        \coordinate (A2) at ($(A) + (0, -\w)$);
        \coordinate (B) at (0.5,\w);
        \coordinate (B2) at ($(B) + (0, -\w)$);
        \coordinate (size) at (\w,-\w);
        % draw boxes and label them
        \draw ($(A)$) rectangle ($(A) + (size)$);
        \draw ($(B)$) rectangle ($(B) + (size)$);
        \node at ($(A) + (\w/2, 6*\hw)$) {$\A$};
        \node at ($(B) + (\w/2, 6*\hw)$) {$\B$};
        % column j
        \draw ($(A) + (\jj-\hw, 0)$) -- ($(A) + (\jj-\hw, -\w)$);
        \draw ($(A) + (\jj+\hw, 0)$) -- ($(A) + (\jj+\hw, -\w)$);
        \draw ($(B) + (\jj-\hw, 0)$) -- ($(B) + (\jj-\hw, -\w)$);
        \draw ($(B) + (\jj+\hw, 0)$) -- ($(B) + (\jj+\hw, -\w)$);
        \node at ($(B) + (\jj, 2*\hw)$) {j};
        \node at ($(A) + (\jj, 2*\hw)$) {j};
        % column j'
        \draw ($(A) + (\jjj-\hw, 0)$) -- ($(A) + (\jjj-\hw, -\w)$);
        \draw ($(A) + (\jjj+\hw, 0)$) -- ($(A) + (\jjj+\hw, -\w)$);
        \draw ($(B) + (\jjj-\hw, 0)$) -- ($(B) + (\jjj-\hw, -\w)$);
        \draw ($(B) + (\jjj+\hw, 0)$) -- ($(B) + (\jjj+\hw, -\w)$);
        \node at ($(B) + (\jjj, 2*\hw)$) {j'};
        \node at ($(A) + (\jjj, 2*\hw)$) {j'};
        % row i
        \draw ($(A) + (0, -\ii-\hw)$) -- ($(A) + (\w, -\ii-\hw)$);
        \draw ($(A) + (0, -\ii+\hw)$) -- ($(A) + (\w, -\ii+\hw)$);
        \draw ($(B) + (0, -\ii-\hw)$) -- ($(B) + (\w, -\ii-\hw)$);
        \draw ($(B) + (0, -\ii+\hw)$) -- ($(B) + (\w, -\ii+\hw)$);
        \node at ($(A) + (-2*\hw, -\ii)$) {i};
        \node at ($(B) + (-2*\hw, -\ii)$) {i};
        % row i'
        \draw ($(A) + (0, -\iii-\hw)$) -- ($(A) + (\w, -\iii-\hw)$);
        \draw ($(A) + (0, -\iii+\hw)$) -- ($(A) + (\w, -\iii+\hw)$);
        \draw ($(B) + (0, -\iii-\hw)$) -- ($(B) + (\w, -\iii-\hw)$);
        \draw ($(B) + (0, -\iii+\hw)$) -- ($(B) + (\w, -\iii+\hw)$);
        \node at ($(A) + (-2*\hw, -\iii)$) {i'};
        \node at ($(B) + (-2*\hw, -\iii)$) {i'};
        % column c
        \draw ($(A) + (\jc-\hw, 0)$) -- ($(A) + (\jc-\hw, -\w)$);
        \draw ($(A) + (\jc+\hw, 0)$) -- ($(A) + (\jc+\hw, -\w)$);
        \draw ($(B) + (\jc-\hw, 0)$) -- ($(B) + (\jc-\hw, -\w)$);
        \draw ($(B) + (\jc+\hw, 0)$) -- ($(B) + (\jc+\hw, -\w)$);
        \node at ($(B) + (\jc, 2*\hw)$) {c};
        \node at ($(A) + (\jc, 2*\hw)$) {c};
        % fill in structural zeros as black
        \draw[fill=black]  
            ($(A2) + (\hw, 0)$) -- 
            ($(A2) + (\hw, 2*\hw)$) -- 
            ($(A2) + (\jjj-\hw, 2*\hw)$) -- 
            ($(A2) + (\jjj-\hw, 4*\hw)$) -- 
            ($(A2) + (\jjj+\hw, 4*\hw)$) -- 
            ($(A2) + (\jjj+\hw, 6*\hw)$) -- 
            ($(A2) + (\jj-\hw, 6*\hw)$) -- 
            ($(A2) + (\jj-\hw, \w-\iii-3*\hw)$) -- 
            ($(A2) + (\jc+\hw, \w-\iii-3*\hw)$) -- 
            ($(A2) + (\jc+\hw, \w-\iii+3*\hw)$) -- 
            ($(A2) + (\w-2*\hw, \w-\iii+3*\hw)$) -- 
            ($(A2) + (\w-2*\hw, \w-\hw)$) -- 
            ($(A2) + (\w, \w-\hw)$) -- 
            ($(A2) + (\w, 0)$) -- 
            cycle;
        \draw[fill=black]  
            ($(B2) + (\hw, 0)$) -- 
            ($(B2) + (\hw, 2*\hw)$) -- 
            ($(B2) + (\jjj-\hw, 2*\hw)$) -- 
            ($(B2) + (\jjj-\hw, 4*\hw)$) -- 
            ($(B2) + (\jjj+\hw, 4*\hw)$) -- 
            ($(B2) + (\jjj+\hw, 6*\hw)$) -- 
            ($(B2) + (\jj-\hw, 6*\hw)$) -- 
            ($(B2) + (\jj-\hw, \w-\iii-3*\hw)$) -- 
            ($(B2) + (\jc+\hw, \w-\iii-3*\hw)$) -- 
            ($(B2) + (\jc+\hw, \w-\iii+3*\hw)$) -- 
            ($(B2) + (\w-2*\hw, \w-\iii+3*\hw)$) -- 
            ($(B2) + (\w-2*\hw, \w-\hw)$) -- 
            ($(B2) + (\w, \w-\hw)$) -- 
            ($(B2) + (\w, 0)$) -- 
            cycle;
        % fill in cells that agree with cross hatch
        \draw[pattern=north west lines]  
            ($(A)$) -- 
            ($(A) + (0, -\ii+\hw)$) -- 
            ($(A) + (\jj+\hw, -\ii+\hw)$) -- 
            ($(A) + (\jj+\hw, -\ii-\hw)$) -- 
            ($(A) + (\w, -\ii-\hw)$) -- 
            ($(A) + (\w, 0)$) -- 
            cycle;
        \draw[pattern=north west lines]  
            ($(B)$) -- 
            ($(B) + (0, -\ii+\hw)$) -- 
            ($(B) + (\jj+\hw, -\ii+\hw)$) -- 
            ($(B) + (\jj+\hw, -\ii-\hw)$) -- 
            ($(B) + (\w, -\ii-\hw)$) -- 
            ($(B) + (\w, 0)$) -- 
            cycle;
        % fill (i', j') with orange
        \draw[fill=orange, line width=0]
            ($(A) + (\jjj-\hw, -\iii+\hw)$) -- 
            ($(A) + (\jjj-\hw, -\iii-\hw)$) -- 
            ($(A) + (\jjj+\hw, -\iii-\hw)$) -- 
            ($(A) + (\jjj+\hw, -\iii+\hw)$) -- 
            cycle;
        \draw[fill=orange, line width=0]
            ($(B) + (\jjj-\hw, -\iii+\hw)$) -- 
            ($(B) + (\jjj-\hw, -\iii-\hw)$) -- 
            ($(B) + (\jjj+\hw, -\iii-\hw)$) -- 
            ($(B) + (\jjj+\hw, -\iii+\hw)$) -- 
            cycle;
        % draw red/blue boundaries around rows being compared
        \draw[color=red, line width=1.5]
            ($(B) + (0, -\ii+\hw)$) -- 
            ($(B) + (0, -\ii-\hw)$) -- 
            ($(B) + (\jc+\hw, -\ii-\hw)$) -- 
            ($(B) + (\jc+\hw, -\ii+\hw)$) -- 
            cycle;
        \draw[color=blue, line width=1.5]
            ($(B) + (0, -\iii+\hw)$) -- 
            ($(B) + (0, -\iii-\hw)$) -- 
            ($(B) + (\jc+\hw, -\iii-\hw)$) -- 
            ($(B) + (\jc+\hw, -\iii+\hw)$) -- 
            cycle;
        \draw[color=red, line width=1.5]
            ($(A) + (0, -\ii+\hw)$) -- 
            ($(A) + (0, -\ii-\hw)$) -- 
            ($(A) + (\jc+\hw, -\ii-\hw)$) -- 
            ($(A) + (\jc+\hw, -\ii+\hw)$) -- 
            cycle;
        \draw[color=blue, line width=1.5]
            ($(A) + (0, -\iii+\hw)$) -- 
            ($(A) + (0, -\iii-\hw)$) -- 
            ($(A) + (\jc+\hw, -\iii-\hw)$) -- 
            ($(A) + (\jc+\hw, -\iii+\hw)$) -- 
            cycle;
        % fill in values
        \node at ($(A) + (\jj, -\ii)$) {0};
        \node at ($(A) + (\jjj, -\ii)$) {1};
        \node at ($(A) + (\jj, -\iii)$) {1};
        \node at ($(B) + (\jj, -\ii)$) {1};
        \node at ($(B) + (\jjj, -\ii)$) {0};
        \node at ($(B) + (\jj, -\iii)$) {0};
    \end{tikscale}
    \caption{Illustration of irreducibility proof in Appendix~\ref{sec:monotonic_proof}. $\A$ and $\B$ have structural zeros (in black), and agree for elements in the hashed region. Their partials sums must agree in the red regions, and their partial sums must agree in the blue regions as well. Four cases are possible for the orange element.}
    \label{fig:irred_proof}
\end{figure}

When structural zeros are present, that is, at least one of $w_{ij}=0$, the Markov chain implied by checkerboard swapping may become reducible. 
Here we prove that in the special case of monotonic structural zeros, irreducibility still holds.
Specifically we show that between any two matrices $\A$ and $\B$, there exists a checkerboard swap which reduces the Hamming distance between them, so that repeated application of such swaps eventually reduces the distance to 0.
%The chain of swaps employed demonstrates the connectedness of the Markov states for $\A$ and $\B$, and since $\A$ and $\B$ are arbitrary, the chain is irreducible.

\textbf{Finding a Swap.}  Consider $\A, \B \in \arcp$, and let $d_H$ be the Hamming distance between them.
The structural zeros are monotonic, so assume the rows and columns have been arranged such that $w_{ij} = 0 \implies w_{kj} = w_{il} = 0$ for $k>i,~l>j$.
This places all the structural zeros in the bottom-right corner (See Figure~\ref{fig:irred_proof}).

Let $(i,j)$ denote the first element where $\A$ and $\B$ differ, where ``first'' means no rows above $j$ contain a difference, and in row $j$, no column to the right of column $i$ contains a difference.
Specifically, $(i,j)$ satisfies:
\begin{enumerate}
    \item $a_{ij} \neq b_{ij}$ 
    \item $a_{kl} = b_{kl}$ for $k<i$ and $1 \leq l \leq n$ ($\A$ and $\B$ agree above row $i$)
    \item $a_{il} = b_{il}$ for $j<l$ ($\A$ and $\B$ agree in row $j$ to the right of column $j$)
\end{enumerate}
In Figure~\ref{fig:irred_proof}, we show agreement between $\A$ and $\B$ with cross hatching.

Without loss of generality, assume $a_{ij} = 0$ and $b_{ij}=1$.
$\A$ and $\B$ must have the same sum in row $i$, but they differ at $(i, j)$, so there must be some column $j'$ such that $a_{ij'} = 1$ and $b_{ij'}=0$.
We also know $j'<j$ because $j$ is the last column where $\A$ and $\B$ differ in row $i$. Similarly, there must be some row $i'$, such that $i'>i$, where $a_{i'j}=1$ and $b_{i'j}=0$, since the column sums agree.
%We also know $i'>i$ because $i$ is the first row where $\A$ and $\B$ differ.

Element $(i',j')$, filled orange in Figure~\ref{fig:irred_proof}, cannot be a structural zero due to the selection criterion %, since that would imply element $(i',j)$ is a structural zero by monotonicity,
so one of the following is true:
\begin{enumerate}
    \item \underline{Checkerboard in $A$ and $B$:} $a_{i'j'} = 0$ and $b_{i'j'} = 1$ so both $\A$ and $\B$ contain a checkerboard with the elements in rows $\{i,i'\}$ and columns $\{j,j'\}$. 
    Performing the swap in either $\A$ or $\B$ reduces $d_H$ by 4.
    \item \underline{Checkerboard in $A$ only:} $a_{i'j'} = b_{i'j'} = 0$. Performing the swap in $\A$ reduces $d_H$ by 2.
    \item \underline{Checkerboard in $B$ only:} $a_{i'j'} = b_{i'j'} = 1$.
    Performing the swap in $\B$ reduces $d_H$ by 2.
    \item \underline{No checkerboard:} $a_{i'j'} = 1$ and $b_{i'j'} = 0$.
\end{enumerate}
Cases (1-3) allow checkerboard swaps which reduce the distance between $\A$ and $\B$ by at least 2. Case (4) does not permit a swap, but if this is the case, a swap can be found as outlined below.

Let $c$ be the last column in row $i'$ which is not a structural zero.  Since row $i'$ in $A$ and row $i'$ in $B$ have the same sum, the \textit{partial} sums up to column $c$ are also equal (see region with blue border in Figure~\ref{fig:irred_proof}). 

%We will show that the partial sums of rows $j$ and $j'$ agree between $\A$ and $\B$, up to some column $c$, then argue that there must be some column $j''<c$ which fulfills the conditions necessary for a swap to be possible.

%Note that row $i'$ in $\A$ and row $i'$ in $\B$ have the same row sums. In this orientation, however, row $i'$ may have structural zeros at the end, so the \emph{partial} sums up to column $c$ also must agree, where $c$ is the last column in row $i'$ which is not a structural zero.  We show the relevant region in Figure $\ref{fig:irred_proof}$ with a blue border.

Note that $j, j' \leq c$. % because otherwise $(i', j)$ and $(i', j')$ must be structural zeros.
In row $i$, $\A$ and $\B$ agree after column $j$ (due to the selection procedure), and since $j<c$, they agree after column $c$ as well.
Therefore the partial sums up to column $c$ must agree in row $i$ as well (see region with red border in Figure~\ref{fig:irred_proof}). 
%We show the relevant region in Figure $\ref{fig:irred_proof}$ with a red border.

We will use the agreement of these partial sums to show that there exists some column $j''$ which permits a checkerboard swap in either $\A$ or $\B$.  Consider the difference matrix $\D = \A - \B$ with elements $d_{kl} \in \{-1, 0, 1\}$.
Because the partial sums described above agree, we have:
\begin{align}
    0 &= \sum_{l=1}^c d_{il} = \sum_{l \in \{1, \ldots, c\}\setminus\{j, j'\}} d_{il},
\end{align}
since $d_{ij} = -1$, $d_{ij'} = 1$ (recall we are in Case 4 from above). Furthermore,
\begin{align}
    0 = \sum_{l=1}^c d_{i'l} \; \implies \; -2  &= \sum_{l \in \{1, \ldots, c\}\setminus\{j, j'\}} d_{i'l},
      %&= 2 + \sum_{l \in \{1, \ldots, c\}\setminus\{j, j'\}} d_{il} 
\end{align}
since $d_{i'j} = d_{i'j'} = 1$.
Therefore there must be some $j'' \in \{1, \ldots, c\}\setminus\{j, j'\}$ such that $d_{ij''}>d_{i'j''}$, and one of the following is true:
\begin{enumerate}
    \item \underline{Checkerboard in B, $d_{ij''}=1$, $d_{i'j''}=0$:} $a_{ij''} = 1$, $b_{ij''} = 0$, and $a_{i'j''} = b_{i'j''} = 1$, a checkerboard swap exists in $\B$ in rows $\{i, i'\}$ and columns $\{j, j''\}$.
    \item \underline{Checkerboard in A, $d_{ij''}=1$, $d_{i'j''}=0$: }$a_{ij''} = 1$, $b_{ij''} = 0$, and $a_{i'j''} = b_{i'j''} = 0$, a checkerboard swap exists in $\A$ in rows $\{i, i'\}$ and columns $\{j, j''\}$.
    \item \underline{Checkerboard in A, $d_{ij''}=0$, $d_{i'j''}=-1$: } $a_{ij''} = b_{ij''} = 1$, and $a_{i'j''} = 0$, $b_{i'j''} = 1$, a checkerboard swap exists in $\A$ in rows $\{i, i'\}$ and columns $\{j, j''\}$.
    \item \underline{Checkerboard in B,$d_{ij''}=0$, $d_{i'j''}=-1$: }$a_{ij''} = b_{ij''} = 0$, and $a_{i'j''} = 0$, $b_{i'j''} = 1$, a checkerboard swap exists in $\B$ in rows $\{i, i'\}$ and columns $\{j, j''\}$.
\end{enumerate}
Thus there exists a checkerboard swap which decreases $d_H$ by 2.
%Since the diameter of the space is bounded, repeated checkerboard swapping will reduce $d$ to 0. The choice of $\A$ and $\B$ is arbitrary, so the space is irreducible.
Weighted curveball trades include checkerboard swaps, so the proof applies to irreducibility of the weighted curveball algorithm as well.
This also implies that the number of swaps between any two matrices is bounded by $\frac{d_H}{2}$, and hence the diameter of the Markov state space is bounded by half the maximum possible Hamming distance between two matrices.

\end{document}